\documentclass[journal]{IEEEtran}
\usepackage[
  left=1.4cm,      % ≥ 1.57 cm
  right=1.4cm,
  top=1.7cm,
  bottom=1.7cm,    % ≥ 4.3 cm
  columnsep=0.25in % ≥ 0.24 in
]{geometry}
\usepackage{amsmath,amssymb}
\usepackage{array}
\usepackage{booktabs}
\usepackage[style=ieee,maxnames=4,minnames=3,maxbibnames=3]{biblatex}

\usepackage{graphicx}
\usepackage[caption=false,font=footnotesize]{subfig} % subfigures for IEEEtran
\usepackage{xcolor}
\usepackage{bm}

\begin{document}

\title{A JEPA-Based Field-Layer World Model for Bridging Channel Prediction and Estimation}
\author{
    Yuzhi~Yang,~
    Brahim~Mefgouda,~
    Hang~Zou,~
    Lina~Bariah,~
    Anis~Bara,
    Yuhuan~Lu,~
    Hao~Zhang,~
    and~M\'erouane~Debbah%\IEEEmembership{Fellow,~IEEE.}
\vspace{-1cm}
\thanks{Y. Yang, B. Mefgouda, H. Zou, L. Bariah, A. Bara, H. Zhang, and M. Debbah are with the Institute of Digital Future, Khalifa University, Abu Dhabi 127788, UAE (e-mails: \{yuzhi.yang, brahim.mefgouda, hang.zou, lina.bariah, anis.bara, hao.zhang, merouane.debbah\}@ku.ac.ae). }
\thanks{Y. Lu is with the Faculty of Applied Sciences, Macao Polytechnic University, Macao SAR, China (e-mail: yhlu@mpu.edu.mo).}
}

\maketitle

\begin{abstract}
Channel state information (CSI) acquisition, reconstruction, and prediction are fundamental yet costly tasks in modern MIMO-OFDM wireless systems. Direct coefficient-level prediction of raw CSI is fragile in realistic propagation environments, since small spatial perturbations, local scattering changes, and phase variations can cause large errors in the complex channel domain. However, the underlying wireless propagation field still contains stable and predictable structures that can be exploited across time, frequency, antenna, and carrier dimensions. Motivated by this observation, we propose a JEPA-based field-layer world model (FWM) that learns a shared latent propagation state from multi-resolution CSI observations across the considered carrier bands and predicts its task-relevant evolution in the latent domain. The proposed FWM maps multiple CSI observation resolutions to a shared latent propagation-field space through scale-specific tokenizer heads. A latent prediction backbone is then trained to infer masked or future field states, while an incremental multi-scale alignment strategy allows new observation scales to be incorporated without retraining the entire model from scratch. 
For downstream reconstruction, the predicted latent field is used as a structured prior and combined with sparse current pilots. Experiments on single-band and cross-band reconstruction demonstrate improved symbol detection and, more notably, substantial beamforming gains despite modest NMSE improvements, indicating that FWM captures task-relevant spatial propagation structure beyond coefficient-wise CSI fitting.
\end{abstract}


\vspace{-0.4cm}
\section{Introduction}
\vspace{-0.2cm}

\subsection{Motivation}

Recent years have witnessed rapid progress in intelligent wireless communications, where learning-based methods are increasingly used to complement conventional signal processing. Among physical-layer problems, channel state information (CSI) acquisition, reconstruction, denoising, and prediction are particularly essential, since accurate CSI is the basis for beamforming, precoding, resource allocation, and other wireless tasks \cite{twm, naoumi2026structuredlatentdynamicswireless}. However, obtaining high-resolution CSI over dense time-frequency resources is costly, especially in wideband, multi-antenna, and mobile scenarios.

As channel prediction can provide proactive channel knowledge and potentially reduce repeated pilot acquisition, it becomes an attractive task to improve the overall system performance. Nevertheless, the prediction target must be carefully defined. Wireless channels are not uniformly predictable in all components. Stable propagation structures are related to scene geometry, blockage, dominant scattering regions, and user mobility, while fine-grained small-scale fading, especially phase-sensitive CSI coefficients, can change sharply under small spatial perturbations, local scattering variations, and hardware-induced phase shifts. As a result, directly minimizing the prediction error in the raw CSI domain may force the model to fit unstable coefficient-level details that are difficult to infer from historical observations. Therefore, raw CSI prediction should not be viewed as a complete replacement for current channel estimation in realistic systems.

This motivates a latent predictive view of wireless channels. Instead of forcing the backbone model to extrapolate all raw CSI coefficients, the model should capture stable and predictable propagation-field structure in a latent representation space. When fine-grained instantaneous details are required, they can be recovered in a downstream stage using available task-side observations, such as sparse pilots. A natural way to realize this principle is the JEPA-style world-model paradigm \cite{ijepa, vjepa, vjepa21, naoumi2026structuredlatentdynamicswireless}, where prediction is performed in latent space rather than in the raw observation domain. This allows the predictive backbone to focus on reusable propagation structure while avoiding direct reconstruction of phase-sensitive details, observation noise, and high-frequency artifacts.

Beyond predictability, practical CSI observations are also highly heterogeneous. Different deployments may use different carrier frequencies, bandwidths, antenna numbers, array geometries, subcarrier spacings, OFDM structures, and pilot patterns. Even for the same propagation environment, the observed CSI tensor may have different dimensions, sampling densities, and sparsity patterns. Existing task-specific models often address channel estimation, reconstruction, or prediction separately under a fixed input format \cite{luo2025ai}. Although large shared models may improve transferability \cite{zhu2025wireless}, increasing model size alone does not specify what physical structure should be shared across tasks and system configurations.

Motivated by these observations, we propose a predictive physical-layer world model, named the field-layer world model (FWM), for heterogeneous wireless channel representation, prediction, and reconstruction. The key idea is to map heterogeneous CSI observations into a shared latent propagation-field space before task-specific processing. Scale- or format-specific tokenizer heads provide an extensible interface for variations in carrier band, antenna configuration, subcarrier spacing, or pilot pattern, while a shared latent backbone models their common propagation semantics. In this way, FWM shifts wireless intelligence from task-specific mapping in the raw channel domain to representation-centric reasoning in a shared physical latent space. This design is intended to support predictable latent-field modeling, heterogeneous observation alignment, and downstream channel reconstruction under imperfect and sparse current observations. The present experiments instantiate this general interface through multi-resolution CSI observations derived from a common full-resolution channel tensor.

\vspace{-0.5cm}
\subsection{Contributions}
\vspace{-0.1cm}
In this paper, we investigate a JEPA-based predictive physical-layer world model for multi-resolution wireless channel observations across the considered carrier bands. Through the JEPA formulation, wireless channels with different scales, times, and frequencies are mapped to the same predictable latent space. With the learned latent representations, we can perform prediction in an implicit domain that is less sensitive to random variations. In particular, the main contributions of this paper can be summarized as follows.
\begin{itemize}
    \item We propose a native AI framework powered by JEPA for wireless systems. The proposed system learns the implicit relationships among the historical channels to generate a latent-field accurate prediction for future channels.
    \item We propose an incremental multi-scale alignment strategy for JEPA training to map different input scales to the same latent space.
    \item We propose applying the trained JEPA results to the reconstruction of the downstream channel, reducing the necessary pilot overhead, and improving the quality of the channel reconstruction.
    \item We further study the effects of different JEPA auxiliary targets and network topologies to derive design principles based on both latent-field prediction and downstream reconstruction quality.
\end{itemize}

\vspace{-0.45cm}
\section{Related Works}
\subsection{Wireless Channel Prediction and Estimation}
In pioneering AI-based channel-domain work, CsiNet showed that massive-MIMO CSI can be compressed into a low-dimensional codeword and recovered by a neural decoder, demonstrating the value of learned channel representations for the acquisition and feedback of CSI \cite{wen2018deep}. Its temporal extension further exploits the correlation among consecutive CSI frames for time-varying massive-MIMO feedback \cite{wang2018deep}. These studies established that neural networks can learn useful channel representations from data, but they mainly focus on CSI compression, feedback, or reconstruction under relatively fixed interfaces, rather than long-horizon prediction of mobile channels or transfer of channel knowledge across carrier frequencies.

Channel prediction has therefore received increasing attention as a way to mitigate channel aging and reduce pilot or feedback overhead in mobile MIMO-OFDM systems. Existing learning-based predictors formulate CSI evolution as a temporal sequence modeling problem and use recurrent networks, convolutional structures, or attention mechanisms to learn correlations from historical channel observations. Representative works include transformer-based channel prediction for mobility-robust CSI forecasting \cite{jiang2022accurate}, spatio-temporal neural networks for massive MIMO-OFDM channel prediction \cite{liu2022spatiotemporal}, aggregated learning for reducing online training overhead in wideband massive MIMO systems \cite{ko2024machine}, and lightweight time-aware transformer architectures for efficient channel prediction \cite{jin2025linformer}.
Beyond purely data-driven sequence modeling, physics-inspired predictors model mobile channel evolution through continuous spatio-temporal dynamics. Neural-ODE-based channel prediction interprets channel variation as a differential process related to user motion and propagation evolution \cite{xiao2023data}, and the more recent ODE-Former further integrates this idea into a former-like architecture to improve prediction accuracy, flexibility, and robustness under non-uniform CSI sampling \cite{xiao2025odeformer}. Recent benchmarking efforts, such as CSI-4CAST, also emphasize robustness and generalization across different channel conditions, mobility regimes, and TDD/FDD settings \cite{cheng2026csi}. However, these methods usually predict channel-domain quantities under a specific CSI format, and do not explicitly separate stable propagation-field structures from phase-sensitive instantaneous coefficient details.

Another closely related direction is cross-frequency channel inference for FDD systems, where uplink and downlink CSI are observed at different carrier frequencies and full coefficient-level reciprocity does not hold. Nevertheless, the two bands may still share propagation geometry, scatterer distribution, delay structure, and angular structure. This motivates uplink-to-downlink channel prediction, frequency extrapolation, and reciprocity-enabling methods \cite{yang2020deep,liu2021fire}.

Recently, large-model-based approaches have further extended channel prediction by treating CSI sequences or channel tensors as structured tokens. LLM4CP adapts pretrained language models for uplink-to-downlink channel prediction \cite{liu2024llm4cp}, and BERT4MIMO explores a BERT-style foundation model for massive-MIMO CSI prediction \cite{catak2025bert4mimo}. In parallel, wireless foundation models such as LWM-Temporal and AirFM-DDA learn transferable representations in angle-delay-time or delay-Doppler-angle domains for channel prediction and estimation \cite{alikhani2026lwm,bian2026airfm}, while recent work formulates CSI prediction in the time-domain and frequency-domain as a unified masked reconstruction problem over heterogeneous space-time-frequency CSI configurations~\cite{liu2025wifo}. These studies suggest that temporal prediction and cross-band transfer are both feasible because wireless channels contain shared physical structures beyond raw CSI coefficients. However, most existing methods still operate on fixed channel formats or directly predict/reconstruct CSI in the signal domain. In contrast, the FWM considered in this paper aligns multi-resolution MIMO-OFDM observations across carrier bands into a shared latent propagation-field space and uses the predicted latent field as a structured prior for sparse-pilot channel reconstruction.
\begin{table*}[t]
    \centering
    \caption{Comparison with representative CSI prediction, wireless foundation-model, and latent predictive-learning methods.}
    \label{tab:method_comparison}
    \setlength{\tabcolsep}{3pt}
    \renewcommand{\arraystretch}{1.25}
    \footnotesize
    \begin{tabular}{
        >{\raggedright\arraybackslash}p{1.9cm}
        >{\raggedright\arraybackslash}p{2.7cm}
        >{\raggedright\arraybackslash}p{2.2cm}
        >{\raggedright\arraybackslash}p{2.2cm}
        >{\raggedright\arraybackslash}p{2.2cm}
        >{\raggedright\arraybackslash}p{2.1cm}
        >{\raggedright\arraybackslash}p{2.9cm}}
        \toprule
        Method &
        Representation / Training Target &
        Heterogeneous CSI Interface &
        Multiple Format Extension &
        Prediction-Estimation Fusion &
        Cross-Frequency Reconstruction &
        Physical-Layer Evaluation \\
        \midrule
        
        ODE-Former~\cite{xiao2025odeformer} &
        Time-domain CSI prediction &
        Fixed format &
        Not demonstrated &
        Not demonstrated &
        Not demonstrated &
        Prediction NMSE and robustness to nonuniform sampling \\

        CSI-4CAST~\cite{cheng2026csi} &
        CSI prediction &
        Multiple scenarios, same CSI scale &
        Not demonstrated &
        Not demonstrated &
        FDD channel-prediction setting &
        Prediction NMSE, robustness, and generalization \\
        
        FIRE~\cite{liu2021fire} &
        Downlink CSI prediction from uplink &
        Multiple datasets with same format &
        Not demonstrated &
        No; prediction is conditioned on uplink CSI &
        Uplink-to-downlink prediction &
        Channel SNR, MIMO SINR, data rate, and runtime \\
        
        \midrule
        
        WiFo~\cite{liu2025wifo} &
        Space-time-freq CSI, masked raw-CSI reconstruction &
        Variable space-time-freq configurations &
        Joint training &
        Not demonstrated &
        Frequency-domain channel extrapolation &
        Time- and freq-domain prediction NMSE \\
        
        LWM-Temporal~\cite{alikhani2026lwm} &
        Angle-delay-time CSI, masked representation learning &
        Multiple datasets with same format &
        Not demonstrated &
        Not demonstrated &
        Not demonstrated &
        Channel-prediction NMSE across mobility regimes \\
        
        AirFM-DDA~\cite{bian2026airfm} &
        Delay-Doppler-angle CSI, masked representation learning &
        Multiple datasets with same format &
        Not demonstrated &
        Separated tasks &
        Not demonstrated &
        Prediction and estimation NMSE \\
        
        \midrule
        
        CSI-JEPA~\cite{luo2026csi} &
        CSI-amplitude windows, masked latent prediction &
        Fixed time-freq observation format &
        Not demonstrated &
        Not demonstrated &
        Not demonstrated &
        Sensing tasks \\
        
        WirelessJEPA~\cite{chu2026wirelessjepa} &
        Multi-antenna I/Q samples; masked latent prediction &
        Fixed time-antenna observation format &
        Not demonstrated &
        Not demonstrated &
        Not demonstrated &
        Classification and regression on multiple RF tasks \\
        
        \midrule
        
        \textbf{Proposed FWM} &
        Multi-resolution CSI across carrier bands, predictive latent propagation field &
        \textbf{Scale-specific tokenizer, shared backbone interface} &
        \textbf{Frozen-reference alignment for newly introduced resolutions} &
        \textbf{History-derived prediction fused with current sparse pilots} &
        \textbf{Observed-to-missing carrier-band reconstruction} &
        \textbf{Reconstruction NMSE, SER, and beamforming gains} \\
        
        \bottomrule
    \end{tabular}

    \vspace{-0.3cm}
\end{table*}
\vspace{-0.4cm}
\subsection{JEPA World Models}
Joint-embedding predictive architectures (JEPAs) learn predictive representations by inferring the latent embeddings of unobserved regions from visible context, rather than reconstructing raw inputs in the observation space \cite{ijepa}. I-JEPA instantiates this principle for
images by predicting target-block embeddings from context-block
embeddings, showing that non-generative latent prediction can learn semantic visual representations without hand-crafted augmentations or contrastive negative samples \cite{ijepa}. V-JEPA extends this idea to videos by predicting masked video features, thereby learning motion-aware and appearance-aware representations without pixel-level reconstruction \cite{vjepa}. More recently, V-JEPA 2 \cite{vjepa2} and V-JEPA 2.1 \cite{vjepa21} further move this line from visual representation learning toward world modeling.
%Meanwhile, other world-model paradigms also learn predictive environment models through generative or model-based objectives. For example, DreamerV3 learns latent dynamics and optimizes behavior by imagination across diverse control domains \cite{hafner2023dreamerv3}, and Genie formulates a generative interactive environment model from unlabeled videos with learned latent actions \cite{bruce2024genie}. 
Compared with these pixel-level or environment-level world models, our FWM adopts the JEPA principle at the wireless field layer, where prediction is performed over stable latent propagation-field representations instead of all raw CSI coefficients.
\vspace{-0.4cm}
\subsection{JEPA in Wireless Communication Systems}
Recently, JEPA-style predictive representation learning has begun to appear in wireless communication systems. A general tutorial on JEPA summarizes its role as a non-generative self-supervised framework that predicts target embeddings in latent space rather than reconstructing raw inputs~\cite{monemi2025tutorial}. Following this idea, a basic idea arises to use JEPA for better channel representation and charting.
Chaaya et al. augment channel charting with a JEPA-style predictor that forecasts future channel embeddings conditioned on user velocity~\cite{chaaya2024learning}. Meanwhile, later work further structure the latent CSI dynamics using Lie-algebra-based transition operators, improving geometric consistency and compositional rollout for channel charting and future embedding prediction~\cite{naoumi2026structuredlatentdynamicswireless}.

Furthermore, WirelessJEPA~\cite{chu2026wirelessjepa} and LatentWave~\cite{mohamed2026latentwave} learn transferable representations from real-world multi-antenna I/Q signals by reshaping them into antenna-time grids and predicting masked latent regions. Their evaluation focuses mainly on downstream RF-centric tasks such as modulation classification, angle-of-arrival estimation, positioning, beam prediction, RF fingerprinting, protocol classification, jamming detection, and interference classification. CSI-JEPA~\cite{luo2026csi} learns temporal-subcarrier CSI representations for WiFi sensing using masked latent prediction and channel-variation-aware masking, and evaluates the frozen encoder on multiple sensing tasks with lightweight adapters.

Another group of works uses world-model ideas to learn multimodal wireless representations. MobiWorld~\cite{chai2025mobiworld} gives a tutorial of a broader mobile-network world model based on controllable generation, where the model provides network element-level observations and system-level feedback for planning and optimization rather than physical-layer channel reconstruction. Telecom world model~\cite{twm} further discusses the usage of world model throughout the whole communication systems. JEPA-MSAC~\cite{zheng2026jepa} maps multimodal sensing and communication measurements into a unified token space and uses temporal block-masked JEPA to support localization, beam prediction, and RSSI prediction. Wireless world model~\cite{chen2026wireless} combines CSI, 3D point clouds, and user trajectories in a JEPA-style multimodal framework, supporting tasks such as CSI prediction, CSI compression and feedback, beam prediction, and localization. TS-JEPA~\cite{girgis2026time} further applies latent predictive modeling to capacity-limited remote control, where embeddings are transmitted and predicted for semantic control and scheduling instead of reconstructing wireless channels.  

Overall, most existing JEPA and wireless world-model studies emphasize representation learning, sensing, beam management, control, channel charting, or network optimization, while only a limited number directly address channel prediction or reconstruction. Table \ref{tab:method_comparison} summarizes the key characteristics and differences of the related works. WiFo is one of the closer works in this direction \cite{liu2025wifo}. However, it follows a reconstruction-oriented foundation-model paradigm rather than a JEPA-style latent predictive prior. Unlike the earlier works, this paper develops FWM for multi-resolution MIMO-OFDM CSI observations across the considered carrier bands, where scale-specific tokenizer heads are aligned into a shared latent propagation-field space. In our considered system, pure channel prediction without any instant channel estimation becomes unrealistic, calling for a new paradigm to integrate prediction with current estimation. The predicted latent field is then used as a prior for sparse-pilot channel reconstruction, including both direct single-band reconstruction and cross-band reconstruction when some current carrier bands have no pilots.

\vspace{-0.3cm}
\section{Field-Layer World Model}

Generally, a FWM is a latent predictive model of the propagation state underlying heterogeneous wireless channel observations. Let \(S_t\) denote the propagation state at time \(t\), and let \(\omega\in\Omega\) denote an observation format determined by factors such as carrier band, antenna configuration, bandwidth, subcarrier sampling, and pilot pattern. A CSI observation can be expressed abstractly as
\begin{equation}
H_t^{(\omega)}
=
\mathcal{O}_{\omega}(S_t)+N_t^{(\omega)},
\end{equation}
where \(\mathcal{O}_{\omega}\) is the format-dependent observation operator and \(N_t^{(\omega)}\) collects measurement noise and other observation-specific perturbations. Thus, CSI tensors with different shapes or sampling patterns may be regarded as different observations of a related underlying propagation state rather than as unrelated signal-domain objects.

The propagation state \(S_t\) is not restricted to a single deterministic parameterization. In a path-based interpretation, a channel at frequency \(f\) may be written as
\begin{equation}
H_t(f)
=
\sum_{p\in\mathcal{P}_t}
\alpha_{p,t}(f)
\mathbf{a}_{p,t}
\exp\left(-\mathrm{j}2\pi f\tau_{p,t}\right),
\end{equation}
where \(\mathcal{P}_t\) is the set of visible propagation paths, \(\alpha_{p,t}(f)\), \(\tau_{p,t}\), and \(\mathbf{a}_{p,t}\) denote the corresponding complex gain, delay, and spatial response. From this perspective, propagation-level structure includes the delay-angle support, dominant paths, path visibility and blockage states, spatial covariance, and dominant spatial eigenspaces. These quantities often evolve more regularly than individual phase-sensitive CSI coefficients and can remain related across observation formats and frequency bands.

In this paper, we use the term ``field layer'' to refer to such a propagation-level structure, and our FWM aims to learn an abstraction on it. It does not denote a direct reconstruction of the continuous electromagnetic field, nor do we assume that each latent coordinate is uniquely identifiable with an individual path, angle, or delay. Instead, the latent field is intended to preserve predictable propagation functionals that are useful across observations and tasks, such as the amplitude structure and the delay-angle power support. Fine-grained instantaneous coefficients that cannot be inferred reliably from history are supplied or corrected by current task-side observations, such as sparse pilots.

Formally, the proposed FWM can be represented by
\begin{equation}
\mathcal{W}_{\rm FWM}
\triangleq
\left(
\{\mathcal{T}_{\omega}\}_{\omega\in\Omega},
\mathcal{F},
\{\mathcal{D}_{q}\}_{q\in\mathcal{Q}}
\right),
\end{equation}
where \(\mathcal{T}_{\omega}\) is a tokenizer for the observation format \(\omega\), \(\mathcal{F}\) is a shared latent prediction backbone and \(\mathcal{D}_{q}\) is a task-specific head for task \(q\). The tokenizer produces a common latent state
\(
Z_t^{(\omega)}
=
\mathcal{T}_{\omega}\left(H_t^{(\omega)}\right),
\)
and the shared backbone predicts missing or future latent states from the visible latent context,
\(
\widehat{Z}_{\mathcal{M}}
=
\mathcal{F}
\left(
Z_{\mathcal{R}},Q_{\mathcal{M}}
\right).
\)
For a downstream task, the corresponding head combines the predicted latent state with any task-side observation \(O_q\), such that
\(
\widehat{Y}_{q}
=
\mathcal{D}_{q}
\left(
\widehat{Z}_{\mathcal{M}},O_q
\right).
\)
For channel reconstruction, \(O_q\) consists of the currently available sparse pilot observations. This formulation separates predictable propagation structure, which is carried by the latent prior, from instantaneous measurement evidence, which is introduced by the downstream task.

Under this definition, the proposed framework satisfies five world-model properties. First, \emph{latent-state abstraction} is provided by the tokenizer bank, which maps heterogeneous CSI observations into a common representation rather than imposing a fixed raw tensor interface. Second, \emph{temporal prediction} is realized by the JEPA backbone, which predicts masked or future field states from historical visible observations. Third, \emph{multi-observation consistency} is encouraged by aligning paired observations of the same underlying channel field across input scales. Fourth, \emph{reusable downstream decoding} is demonstrated by combining the same type of predicted latent prior with task-side observations for single-band reconstruction, cross-band reconstruction, symbol detection, and beamforming evaluation. Fifth, \emph{adaptation to a new observation format} is supported by adding and aligning a new tokenizer while retaining the previously learned latent backbone. The experiments in this paper validate the last property for observation-scale expansion; adaptation across arbitrary carrier bands, array geometries, and pilot formats remains to be evaluated.

Accordingly, the proposed FWM is a predictive physical-layer world model that organizes heterogeneous CSI observations around a shared latent representation of the propagation environment. In this formulation, world modeling is instantiated at the physical-layer perception-prediction interface, where the tokenizer bank defines a common latent propagation state, the shared JEPA backbone models its evolution and conditional inference from partial or historical observations, and the downstream heads translate the predicted state into task-relevant channel information. The central modeled object is therefore the latent propagation state and its evolution, rather than a task-specific mapping between fixed CSI tensors. Through this common state interface, the predicted latent field can be transferred across observation scales and combined with current measurement evidence to support channel reconstruction, symbol detection, and beamforming.

\vspace{-0.35cm}
\section{JEPA Tasks and Training Targets}
\vspace{-0.1cm}
In this section, we describe the JEPA pretraining stage of FWM. The tokenizer is trained to map the original channel observations into a shared latent space through a prediction task. Because latent mapping and prediction are learned jointly, we use a gradual schedule that increases the prediction difficulty over time. We also introduce the different input scales progressively, from small to large.

\begin{figure}[t]
    \centering
    \includegraphics[width=0.85\linewidth]{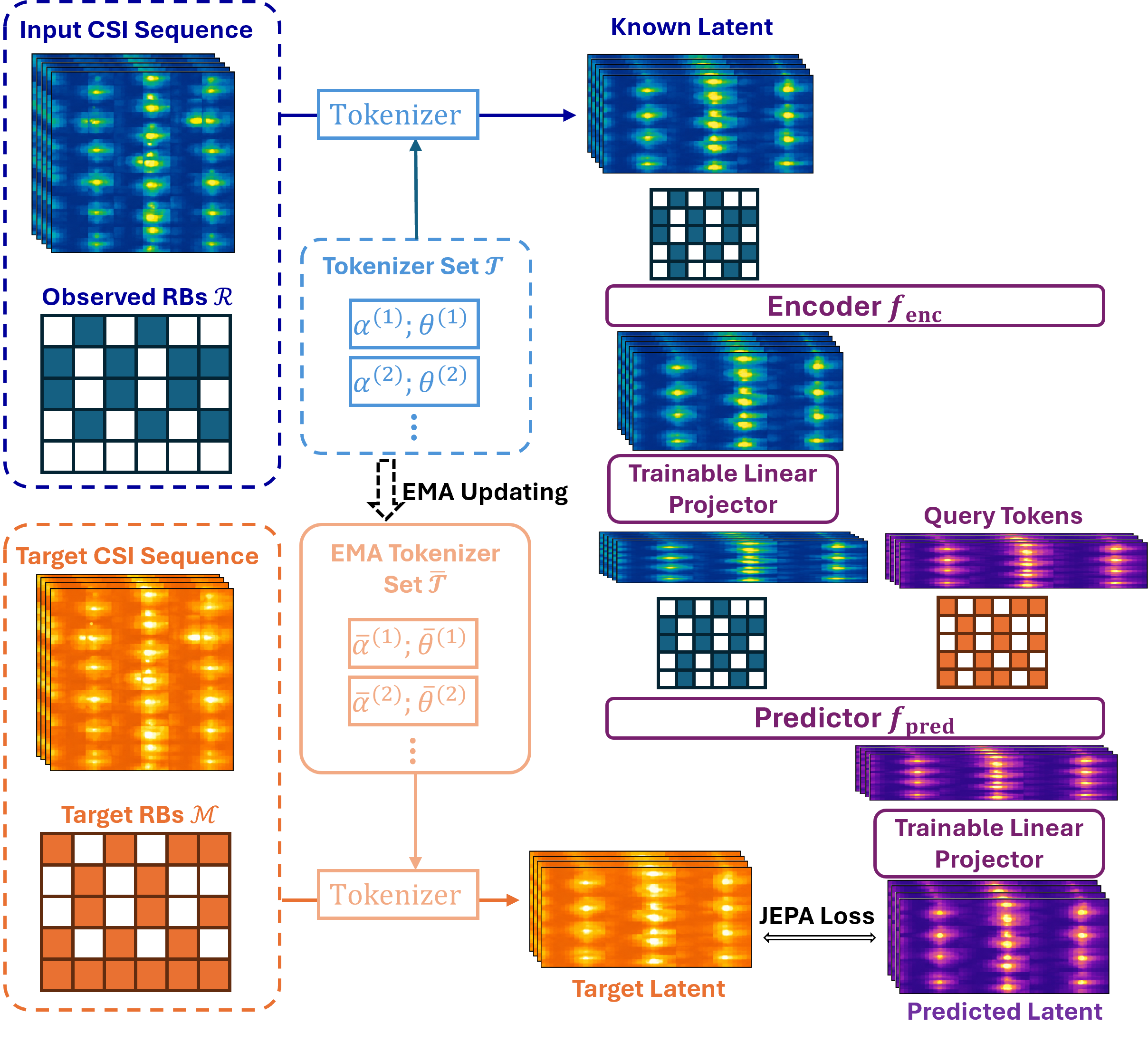}
\vspace{-0.4cm}
    \caption{Overview of the JEPA channel representation and prediction architecture.}
    \label{fig:JEPA}
\vspace{-0.5cm}
\end{figure}
\vspace{-0.35cm}
\subsection{JEPA-Based Latent Field Tokenization and Prediction}

In the first JEPA-based training stage, we adopt the overall scheme shown in Fig.~\ref{fig:JEPA}. This stage aims to map multi-resolution CSI observations into a shared latent field and to train the backbone to predict missing or future latent field states from visible context.

Let \(H_{n,t}\) denote the CSI observation at the carrier band indexed by \(n\) and time slot \(t\). The observation may be complete, partially observed, or represented under a specific system configuration. We use \(\omega_{n,t}\in\mathcal{W}\) to denote its observation format, such as carrier band, antenna configuration, spatial resolution, subcarrier spacing, or pilot pattern. The tokenizer bank is defined as
\(
\mathcal{T}_{\theta}
=
\{\alpha^{(\omega)}(\cdot;\theta^{(\omega)})\,|\,\omega\in\mathcal{W}\}.
\)
For a visible cell \((n,t)\in\mathcal{R}\), the corresponding tokenizer head maps the CSI observation into \(K\) latent tokens of dimension \(D\):
\begin{equation}
L_{n,t}
=
\alpha^{(\omega_{n,t})}
\left(
H_{n,t};\theta^{(\omega_{n,t})}
\right)
\in
\mathbb{R}^{K\times D}.
\end{equation}
Here, \(\mathcal{R}\) denotes the set of visible cells, \(\mathcal{A}\) denotes the set of all cells, and \(\mathcal{M}=\mathcal{A}\setminus\mathcal{R}\) denotes the masked cells. The notation \(\alpha^{(\omega_{n,t})}\) denotes the tokenizer selected from the tokenizer bank according to the input format. The detailed architectures of the tokenizer heads are given in Appendix~\ref{app:tokenizer}.

The tokenized outputs form a latent time--carrier-band field. Each cell is associated with its global carrier-band and time indices \((n,t)\), while the \(K\) tokens inside the cell represent local latent components of that CSI observation. The time and carrier-band indices are encoded by two-dimensional positional encoding in the attention layers, so that multi-resolution CSI observations are converted into compatible latent field tokens with shared spatial, spectral, and temporal semantics.

The visible latent field is denoted by
\(
\mathcal{L}_{\mathcal{R}}
=
\{L_{n,t}\mid (n,t)\in\mathcal{R}\}.
\)
A context encoder processes only the visible latent cells and produces contextualized latent representations
\(
\mathcal{Z}_{\mathcal{R}}
=
f_{\rm enc}
\left(
\mathcal{L}_{\mathcal{R}};\phi_{\rm enc}
\right).
\)
The masked cells are not provided to the context encoder. Therefore, the encoder is forced to summarize the available time--carrier-band context in the shared latent field.

A predictor network then infers the latent states of the masked cells from the contextualized visible field. For each masked cell \((n,t)\in\mathcal{M}\), a learnable query token group \(Q_{n,t}\in\mathbb{R}^{K\times D_p}\) is introduced to specify the target position. The predictor takes the visible context and the target queries as input and outputs the predicted masked latents:
\begin{equation}
\widehat{\mathcal{L}}_{\mathcal{M}}
=
f_{\rm pred}
\left(
\mathcal{Z}_{\mathcal{R}},
\{Q_{n,t}\}_{(n,t)\in\mathcal{M}};
\phi_{\rm pred}
\right),
\end{equation}
where
\(
\widehat{\mathcal{L}}_{\mathcal{M}}
=
\{\hat{L}_{n,t}\in\mathbb{R}^{K\times D}\mid (n,t)\in\mathcal{M}\}.
\)

The target latents for the masked cells are generated by the target tokenizer bank, which is introduced in the next subsection. The detailed structures of the context encoder, predictor, and projection layers are provided in Appendix~\ref{app:pred}.
\vspace{-0.4cm}
\subsection{JEPA Loss}

We first consider the JEPA training objective after the CSI observations have been mapped into the shared latent field. The online branch uses the online tokenizer bank and the context encoder to process visible cells, while the target latents of masked cells are produced by an exponential-moving-average (EMA) target tokenizer bank.

Let
\(
\bar{\mathcal{T}}_{\bar{\theta}}
=
\{
\bar{\alpha}^{(\omega)}(\cdot;\bar{\theta}^{(\omega)})
\,|\, \omega\in\mathcal{W}
\}
\)
denote the target tokenizer bank, which has the same architecture as the online tokenizer bank but uses EMA-updated parameters. For each masked cell \((n,t)\in\mathcal{M}\), the target latent is obtained from the tokenizer head corresponding to its input format
\begin{equation}
L_{n,t}^{\rm tar}
=
\bar{\alpha}^{(\omega_{n,t})}
\left(
H_{n,t};
\bar{\theta}^{(\omega_{n,t})}
\right),
\quad
(n,t)\in\mathcal{M}.
\end{equation}
The target branch is stopped from gradient back-propagation and only provides stable prediction targets. After each training step, the target tokenizer parameters are updated by EMA:
\begin{equation}
\bar{\theta}^{(\omega)}
\leftarrow
\mu\bar{\theta}^{(\omega)}
+
(1-\mu)\theta^{(\omega)},
\quad
\omega\in\mathcal{W},
\end{equation}
where \(\mu\) is the momentum coefficient.

The main JEPA objective is the masked latent prediction loss
\begin{equation}
\ell_{\rm JEPA}
=
\frac{1}{|\mathcal{M}|}
\sum_{(n,t)\in\mathcal{M}}
\frac{1}{KD}
\left\|
\hat{L}_{n,t}
-
L_{n,t}^{\rm tar}
\right\|_{1},
\end{equation}
If variable-length trajectories are used, the summation is taken only over valid masked cells. Since a prediction loss alone may admit collapsed or overly simplified latent representations, we further introduce the regularization terms described in the next subsection.

\vspace{-0.6cm}
\subsection{Diversity Regularization and Grounding Ablations}
\vspace{-0.15cm}

The JEPA prediction loss constrains the predicted masked latents, but it does not explicitly encourage the latent field to be diverse across different samples. To reduce redundant latent states, we introduce a cross-sample diversity loss on the visible-cell latents. Let \(l_i=\operatorname{vec}(L_i)\in\mathbb{R}^{KD}\) denote the flattened latent tokens of the \(i\)-th visible cell in a mini-batch, and let \(B\) be the number of such cells. After centering the rows, we stack them into \(L\in\mathbb{R}^{B\times KD}\) and define the sample-similarity matrix
\(
    S = \frac{1}{KD}LL^{T}.
\)
The diversity loss is then defined as
\begin{equation}
    \ell_{\rm div}
    =
    \frac{1}{B(B-1)}
    \sum_{i\neq j} S_{ij}^{2}.
\end{equation}
This loss discourages latent states from different samples from becoming overly similar. We use a cross-sample form because latent tokens within the same trajectory or propagation path may be naturally correlated and should not be forced to be decorrelated.

In addition to the main JEPA-diversity objective, we also evaluate several CSI-domain grounding objectives as ablations. These objectives share the generic form
\begin{equation}
    \ell_{\rm g}
    =
    \frac{1}{|\mathcal{M}|}
    \sum_{(n,t)\in\mathcal{M}}
    d
    \left(
        D_{\rm g}(U^{\rm g}_{n,t}),
        V^{\rm g}_{n,t}
    \right),
\end{equation}
where \(D_{\rm g}\) is a lightweight grounding head, \(U^{\rm g}_{n,t}\) and \(V^{\rm g}_{n,t}\) denote the input and target of a specific grounding variant, and \(d(\cdot,\cdot)\) is the corresponding regression loss. In our ablations, direct CSI recovery uses the latent code to recover the normalized CSI, amplitude recovery replaces the target by the channel amplitude, and diffusion grounding uses the latent code together with a noisy CSI sample and diffusion step to predict the injected noise. The detailed input-target definitions and grounding-head architectures are provided in Appendix~\ref{app:aux}.

The default single-scale JEPA objective is
\begin{equation}
    \ell_{\rm single}
    =
    \ell_{\rm JEPA}
    +
    \lambda_{\rm div}\ell_{\rm div}.
\end{equation}
When CSI-domain grounding is enabled for ablation studies, the objective becomes
\begin{equation}
    \ell_{\rm single}^{\rm abl}
    =
    \ell_{\rm JEPA}
    +
    \lambda_{\rm div}\ell_{\rm div}
    +
    \lambda_{\rm g}\ell_{\rm g}.
\end{equation}
Only one grounding variant is enabled in each ablation setting.
\vspace{-0.85cm}
\subsection{Multi-Scale Alignment}
\vspace{-0.1cm}

The single-scale JEPA objective can be directly extended to multi-scale inputs by assigning each scale its own tokenizer and applying the same latent prediction loss. This joint training strategy is valid in principle. However, it requires the model to simultaneously learn suitable tokenizer heads for different input scales, align the latent spaces produced by these tokenizers, and learn predictive dynamics in the shared latent field. These coupled objectives may impose a heavy optimization burden, especially when new scales correspond to different resource allocation patterns or sampling resolutions.

We therefore introduce an incremental multi-scale alignment strategy. Suppose that an FWM has already been trained on a reference scale. A new tokenizer head is then introduced for a target scale, while the reference scale may also have a corresponding student branch in the mixed-scale model. This setting naturally arises when time-frequency resources are not fully allocated to a single user, so that different observations of the same underlying channel field may correspond to different sampling patterns or resolutions. Here we assume that the old reference scale can be downsampled from the new target one, and both scales are available everywhere in the dataset.\footnote{This paired construction is a simplified setup for the staged alignment experiment, not a requirement of the framework. In practical datasets, new-scale observations may be available only for part of the samples. In that case, the alignment loss can be applied only to available cross-scale pairs, while unpaired samples can still be used for the student-side JEPA prediction objective or skipped in the alignment term.}

We keep the same visible/masked cell partition as in the single-scale JEPA task. The visible set \(\mathcal{R}\) is further divided into
\(\mathcal{R}
=
\mathcal{R}_{\rm old}
\cup
\mathcal{R}_{\rm new},\)
where \(\mathcal{R}_{\rm old}\) contains visible cells represented by an already learned scale, and \(\mathcal{R}_{\rm new}\) contains visible cells represented by the newly introduced scale. The context encoder receives the mixed-scale visible latents from the whole \(\mathcal{R}\), while the cells in \(\mathcal{M}\) are masked from the context encoder and predicted by the JEPA predictor. This keeps the prediction logic consistent with the single-scale formulation, while allowing the visible context itself to contain multiple input scales.

The alignment objective is applied to all paired cells rather than to a separate set of alignment cells. For each \((n,t)\in\mathcal{A}\), the reference branch produces a latent \(L_{n,t}^{\rm ref}\), and the student branch produces the corresponding latent \(L_{n,t}^{\rm stu}\) from the assigned input scale. The alignment loss is
\begin{equation}
\ell_{\rm align}
=
\frac{1}{|\mathcal{A}|}
\sum_{(n,t)\in\mathcal{A}}
\frac{1}{KD}
\left\|
L_{n,t}^{\rm stu}
-
L_{n,t}^{\rm ref}
\right\|_2^2 .
\end{equation}
For masked cells, the student latents used in this alignment term are generated only for supervision and are not provided to the context encoder. Thus, the model still performs masked latent prediction in the same way as the single-scale JEPA task.

Alignment alone is not sufficient, because the mixed-scale student model must also learn to predict masked latent states from mixed-scale visible context. Therefore, we retain the student-side JEPA prediction loss on the masked set \(\mathcal{M}\), denoted by \(\ell_{\rm single}^{\rm stu}\), following the same form as the single-scale loss. The incremental multi-scale alignment objective is
\begin{equation}
\ell_{\rm MSA}
=
\lambda_{\rm align}(e)\ell_{\rm align}
+
(1-\lambda_{\rm align}(e))\ell_{\rm single}^{\rm stu}.
\end{equation}
In practice, the weights can be scheduled during training. The alignment term is emphasized in the early stage so that the new tokenizer head enters the pretrained latent field, while the JEPA term is strengthened later to refine mixed-scale latent prediction.

The update rule of the reference-scale branch is left as an experimental choice. It can be frozen, updated by EMA, or partially updated together with the student branch. These variants are compared in the experimental section.

\begin{figure}[t] \centering \includegraphics[width=0.8\linewidth]{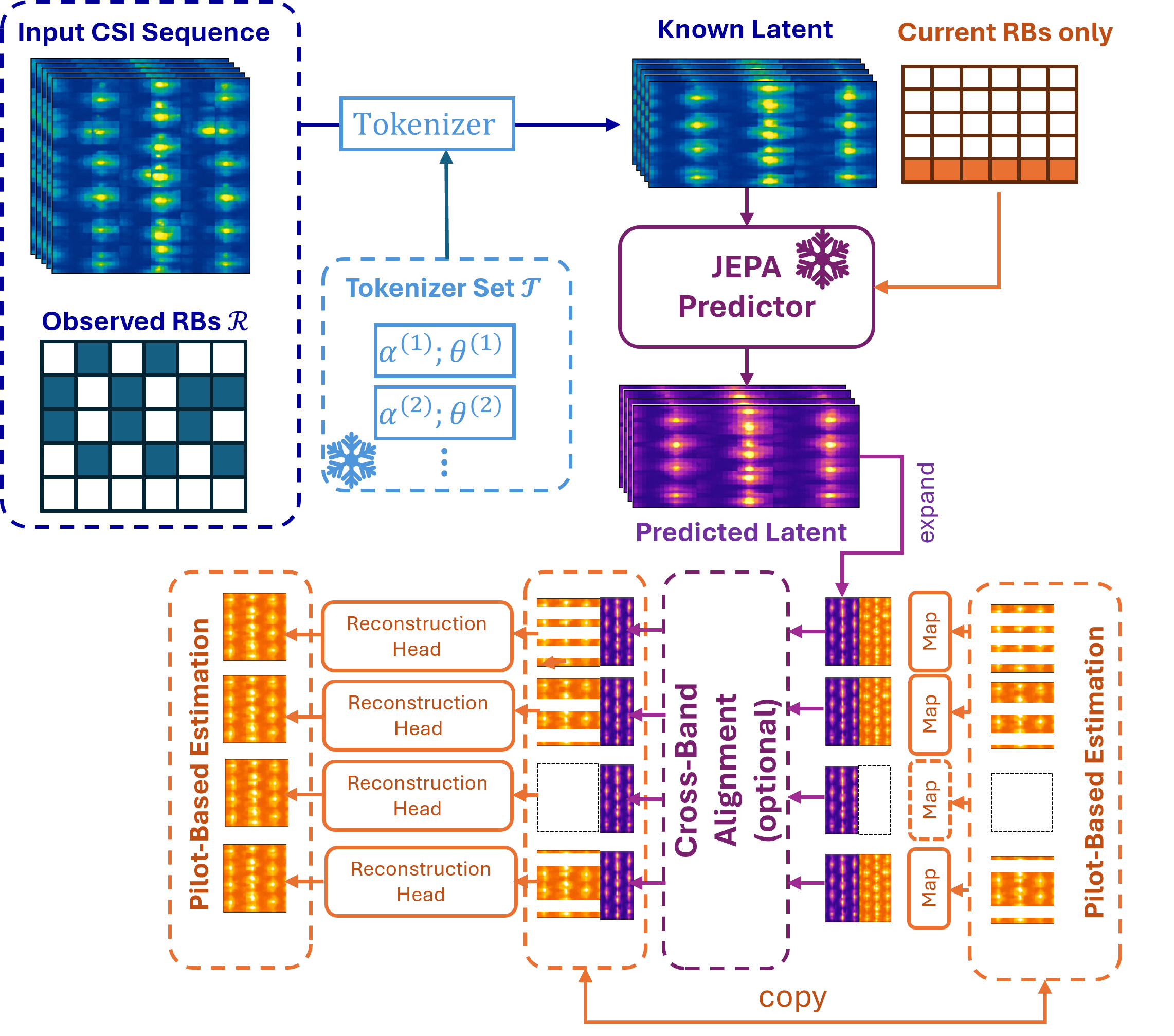} \caption{Illustration of the downstream channel estimation task. The cross-band alignment part may be omitted in some experiments.} \label{fig:CE}
\vspace{-0.55cm} \end{figure}
\vspace{-0.5cm}
\section{Downstream Channel Reconstruction}
\vspace{-0.1cm}

After the JEPA-based FWM pretraining stage, the learned latent field can be used as a propagation-field prior for channel-domain reconstruction tasks. The key principle is that the FWM is not expected to replace the current channel observations. Instead, it provides a structured latent prior learned from historical and cross-band contexts, while sparse current pilots provide instantaneous measurement evidence for recovering fine-grained channel details.

In this section, we consider a downstream channel reconstruction problem. The target is to reconstruct the current MIMO-OFDM channel from two sources of information: the latent field predicted by the pretrained FWM and the currently available pilot-based channel observations. Depending on pilot availability, we consider two downstream settings. The first is direct reconstruction, where each target carrier band has sparse pilot observations and can be reconstructed independently using the latent prior.
We then consider cross-band alignment, where we need to combine the predicted latent field with pilot information from observed carrier bands to enhance the reconstruction of other carrier bands. We note that the direct prediction method without current estimation usually cannot work well due to the random disturbances in the scenario. However, the assistance of other carrier bands may facilitate the observation of phase-sensitive parameters, which is widely observed in the FDD dual link mapping problem \cite{liu2021fire}.

\vspace{-0.5cm}
\subsection{Direct Single-Band Channel Reconstruction}

We first consider the basic setting in which each carrier band contains a local pilot-based estimate, which may be inaccurate. For each carrier band \(n\) at the current time slot \(T\), the sparse pilot observation is modeled as
\begin{equation}
Y_{n,T}
=
M^{\rm p}_{n,T}
\odot
\left(
H_{n,T}
+
W_{n,T}
\right),
\end{equation}
where \(M^{\rm p}_{n,T}\) is a binary pilot mask indicating the positions of sparse pilots, \(W_{n,T}\) denotes the original estimation noise, and \(\odot\) is element-wise multiplication. This formulation ensures that noise is applied only to observed pilot positions.

The pretrained FWM predicts a latent prior \(\hat{L}_{n,T}\) for the current carrier band from historical visible observations. This latent prior captures predictable propagation-field structure, but it does not necessarily contain all instantaneous channel details. Therefore, it is combined with the sparse current pilot observation by a supervised channel reconstruction head:
\begin{equation}\label{down}
\hat{H}_{n,T}
=
h_{\rm rec}
\left(
Y_{n,T},
\hat{L}_{n,T};
\psi_{\rm rec}
\right).
\end{equation}
Here, \(h_{\rm rec}\) denotes the single-band reconstruction head. The head directly learns to fuse the predicted latent prior with the sparse pilot evidence.

The training loss is defined as
\begin{equation}\label{loss_down}
\ell_{\rm rec}
=
\mathbb{E}
\left\|
\hat{H}_{n,T}-H_{n,T}
\right\|_2^2
.
\end{equation}
This task corresponds to the standard sparse-pilot channel reconstruction setting. The predicted latent prior provides historical propagation-field information, while the current pilots correct instantaneous fine-grained channel details.

\vspace{-0.45cm}
\subsection{Cross-Band Alignment}
The direct single-band setting assumes that each target carrier band has its own sparse pilot observations. However, in practical systems, current pilots may be available only in a subset of the carrier bands. The remaining carrier bands then have no direct pilot observations. A related case arises in FDD systems, where the BS needs to infer the downlink channel from uplink transmissions. In this case, the missing carrier bands cannot be reconstructed by local pilot evidence alone.

To handle this setting, we introduce cross-band alignment. Let \(\mathcal{N}_{\rm obs}\) denote the set of carrier bands with current pilot observations, and let \(\mathcal{N}\backslash \mathcal{N}_{\rm obs}\) denote the set of carrier bands without current pilots. For \(n\in\mathcal{N}_{\rm obs}\), the pilot observation \(Y_{n,T}\) and pilot mask \(M^{\rm p}_{n,T}\) are available. For \(n\notin\mathcal{N}_{\rm obs}\), no current pilot observation is available, and the model must rely on the predicted latent prior together with information transferred from the observed carrier bands.

For the observed carrier bands, a pilot-aware tokenizer maps the sparse pilot observation into the shared latent space:
\begin{equation}
L^{\rm p}_{n,T}
=
\alpha^{\rm p}
\left(
Y_{n,T};
\theta^{\rm p}
\right),
\quad n\in\mathcal{N}_{\rm obs}.
\end{equation}
Here we note that it is not the same as the JEPA tokenizer. Instead, this pilot-conditioned latent contains instantaneous measurement evidence from the current time slot. Thus, the working latent space of $L^{\rm p}_{n,T}$ also differs from that of $L_{n,T}$ in the JEPA task.

The cross-band calibration module then refines the predicted latent priors using the available pilot-conditioned latents
\begin{equation}
\{
\bar{L}_{k,T}
\}_{k\in\mathcal{N}}
=
\Gamma_{\rm cali}
\left(
\{\hat{L}_{k,T}\}_{k\in\mathcal{N}},
\{L^{\rm p}_{n,T}\}_{n\in\mathcal{N}_{\rm obs}};
\psi_{\rm cali}
\right).
\end{equation}
For observed carrier bands, the refined latent \(\bar{L}_{n,T}\) incorporates both the prior prediction of the propagation-field and current pilot evidence. For missing carrier bands, \(\bar{L}_{m,T}\) is inferred from the predicted latent prior and cross-band information from the observed ones.

Finally, a supervised channel reconstruction head reconstructs each carrier band according to \eqref{down}. For \(k\in\mathcal{N}_{\rm miss}\), \(Y_{k,T}\) and \(M^{\rm p}_{k,T}\) are set to zero inputs. Thus, missing carrier bands are reconstructed from the calibrated latent prior rather than from local pilot observations.
The same training loss in \eqref{loss_down} is also applied to both observed and missing carrier bands.

This task evaluates whether the FWM can reconstruct channel responses when some current carrier bands have no direct pilots. The observed carrier bands provide current measurement evidence, while the latent field transfers related propagation information to the missing carrier bands.
\vspace{-0.4cm}
\section{Simulation Setup and Common Settings}
\vspace{-0.1cm}
\begin{figure}[t]
    \centering
    \includegraphics[width=0.75\linewidth]{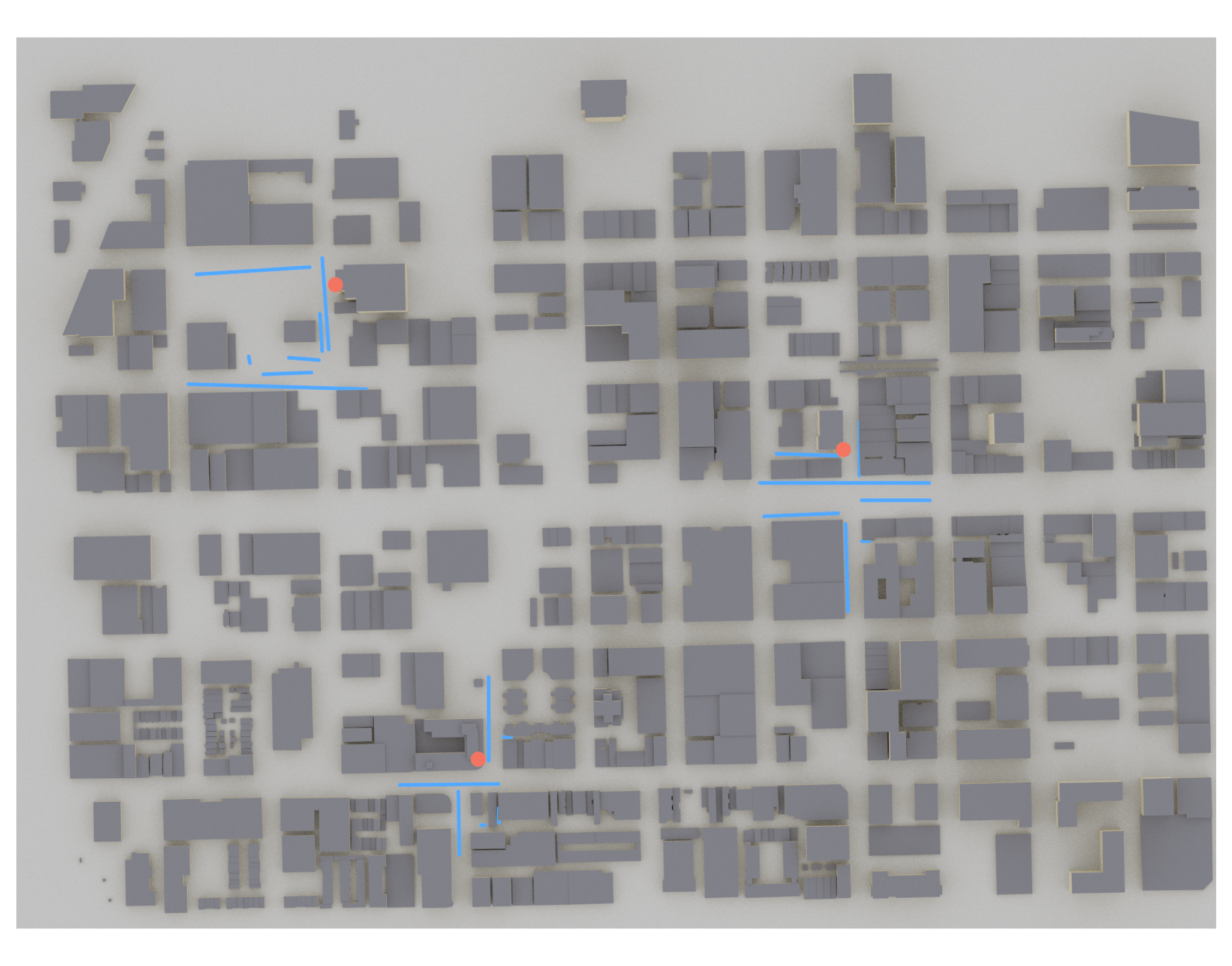}
    \caption{The scenario considered for generating dataset based on the actual maps in Chicago within a range of around 1 km$^2$. The red point indicates the BSs and blue lines are UE paths.}
    \label{fig:scenario}
\end{figure}
\subsection{Dataset Generation}
We generate the channel dataset using Sionna RT over the city-center scenario of Chicago as illustrated in Fig. \ref{fig:scenario}. Three BSs are placed on building rooftops, and 200 UE trajectories are generated for each BS. For each trajectory, the starting and ending points are randomly placed on the ground within a horizontal distance of 100 m from the corresponding BS, with a trajectory length between 1 m and 10 m. Each trajectory contains 16 time samples that are approximately uniformly distributed along the path, and the time interval between adjacent samples is 50 ms. To avoid unrealistically regular phase evolution along perfectly smooth trajectories, each UE position is further perturbed by a Gaussian spatial shift with standard deviation 5 cm. Complete UE trajectories are first split into disjoint training and testing sets at the trajectory level, and all temporal slicing and observation construction are then performed independently within each split.\footnote{All dataset and codes of this paper can be found at \url{https://github.com/yuzhiyang123/FWM_JSAC}}

The ray-tracing simulator generates high-resolution channel matrices of size \(1024\times64\) for each time--carrier-band cell, where the two dimensions correspond to OFDM subcarriers and BS antenna elements, respectively. The 64 antenna elements form an \(8\times8\) array. We consider five sub-\(6\)-GHz carrier bands at 3.1, 3.3, 3.5, 3.7, and 3.9 GHz with original subcarrier spacing 15 kHz, and three mmWave carrier bands at 27, 28, and 29 GHz with original subcarrier spacing 240 kHz. Therefore, each full channel trajectory contains 16 time samples, 8 carrier bands, 1024 OFDM subcarriers per band, and 64 antenna elements.

More specifically, the propagation paths are first computed by Sionna RT at \(3.5\) GHz using a common scene geometry and material configuration. The resulting path gains, delays, and angles of arrival and departure are then reused to synthesize the channel responses at all considered carrier frequencies, while the carrier-dependent phase, Doppler, array response, and OFDM frequency response are recomputed for each band. Consequently, the present dataset preserves shared geometric propagation structure across the sub-\(6\)-GHz and millimeter-wave bands, but does not model frequency-dependent material responses or band-dependent path visibility, such as paths appearing or disappearing because of frequency-selective penetration, reflection, diffraction, or blockage. The cross-band experiments should therefore be interpreted as evaluating representation transfer under a shared-path geometric model.

Different observation scales used in the experiments are constructed from the high-resolution channel matrices during training. For the \(128\times64\) scale, all 64 antenna elements are kept, and one OFDM subcarrier is uniformly sampled from every group of 8 OFDM subcarriers. For the \(32\times16\) scale, one OFDM subcarrier is uniformly sampled from every group of 32 OFDM subcarriers, and the antenna dimension is reduced by selecting one \(4\times4\) subarray after evenly partitioning the original \(8\times8\) array into four non-overlapping regions. The starting OFDM-subcarrier offset is randomly sampled during training and kept fixed along each trajectory. This offset is not stored as a fixed part of the dataset, but is sampled as part of the observation construction process.

\subsection{Phase Sensitivity to Position Uncertainty}
Under the standard multipath channel model, a UE displacement \(\delta\mathbf{r}\) changes the phase of path \(\ell\) by approximately
\(
\delta\phi_{\ell}
\simeq
-\frac{2\pi}{\lambda}
\mathbf{u}_{\ell}^{T}\delta\mathbf{r},
\)
where \(\mathbf{u}_{\ell}\) is its propagation direction. With the \(50\)-ms sampling interval used in our dataset, a \(5\)-cm unmodeled displacement corresponds to only \(1~\mathrm{m/s}\) of velocity uncertainty. Nevertheless, \(5\) cm equals \(0.58\lambda\) at \(3.5\) GHz and \(4.67\lambda\) at \(28\) GHz, producing phase uncertainties of \(3.67\) rad (\(210^\circ\)) and \(29.34\) rad (\(1681^\circ\)), respectively.

Even for a single path with perfectly predicted amplitude, the normalized coefficient error is
\(
\frac{|\hat{h}-h|^2}{|h|^2}
=
4\sin^2\left(\frac{\delta\phi}{2}\right),
\)
and can therefore remain large solely because of phase uncertainty. Moreover, because the position perturbations are independent across adjacent samples, their phase correlation is
\begin{equation}
\rho_{\rm adj}
=
\exp\left[
-\left(\frac{2\pi\sigma_p}{\lambda}\right)^2
\right],
\end{equation}
which is almost  zero at both 3.5 and 28 GHz for \(\sigma_p=5\) cm. Thus, a realistic mobility uncertainty can destroy coefficient-level phase predictability and raw-CSI MSE, while slower propagation structures such as amplitude distributions, angular-delay support, spatial covariance, and dominant eigenspaces may remain predictable.
\vspace{-0.4cm}
\subsection{Training Settings}
Unless otherwise stated, JEPA pretraining uses history masking over the time--carrier-band latent field. Random masking is applied only to historical time steps, while the final time step is always fully masked so that the predictor must infer the target latent field entirely from preceding observations.

For models trained from scratch, we use a progressive masking strategy. At the beginning of training, a fixed historical masking probability is used to stabilize the tokenizer and latent-field learning. The masking probability is then gradually shifted to the default stochastic distribution. Specifically, for epoch \(e\), the historical masking probability is
\begin{equation}
      p_{\rm mask}^{(e)}
      =
      (1-\lambda_e)\cdot 0.3
      +
      \lambda_e\cdot (0.1+0.8u),
      \;
      u\sim{\rm Beta}(4.0,1.2),
\end{equation}
where \(\lambda_e\) increases linearly from 0 to 1 during training. This schedule progressively increases the diversity and difficulty of the masked prediction task while keeping the final time step fully masked.

For training stages initialized from an existing pretrained FWM, such as incremental scale alignment, we directly use the final masking distribution without the progressive warm-up, i.e., $\lambda_e$ is always 1. In these cases, the reference latent field has already been learned, and the training focuses on adapting or aligning the new tokenizer or prediction branch under the default visibility setting.

For downstream channel reconstruction, the model always operates on the full \(1024\times64\) channel domain. We use \(P\) to denote the number of pilot-bearing OFDM subcarriers available in each observed carrier band. The corresponding pilot locations are selected with a random starting point and approximately uniform spacing, so a smaller \(P\) represents lower instantaneous pilot overhead. Additive noise is applied only to the selected pilot entries, and the noise level is specified by the resulting pilot-domain NMSE. The downstream tasks therefore evaluate whether the pretrained latent field, together with \(P\) sparse noisy pilot observations, can recover the full-resolution channel response. Unless otherwise stated, all model architectures are given in the appendix, while the main text focuses on the training objectives, experimental settings, and reconstruction performance.

All experiments use batch size 32 and weight decay \(10^{-4}\). All JEPA training from scratch is conducted for 1000 epochs, and incremental training for each added scale uses 500 epochs. Each downstream model is trained for 200 epochs, except DT, which is trained for 500 epochs because learning the complete predictor--reconstructor from scratch is more difficult. In all cases, the allotted training schedule is sufficient for convergence. The learning rate is linearly warmed up to \(10^{-4}\) during the first 20\% of training and then decayed by a cosine scheduler. For all EMA target updates used in FWM pretraining and alignment, the momentum coefficient is fixed as \(\mu=0.95\). The detailed architectures of all neural-network components are shown in the appendices.
\vspace{-0.3cm}
\section{Latent Field Pretraining Results}
\label{sec:pretraining-results}
\vspace{-0.05cm}

In this section, we evaluate whether the proposed FWM can learn a stable and predictable latent field before discussing the downstream channel reconstruction task. We first select the single-scale objective on \(32\times16\) observations, then compare direct and incremental training across observation scales, and finally study the effect of latent and backbone capacity under a fixed staged training strategy. All results in this section focus on latent prediction and cross-scale alignment, while their downstream effects are discussed below in Section~\ref{sec:downstream-results}.
\begin{figure}[t]
  \centering
  \includegraphics[width=0.65\linewidth]{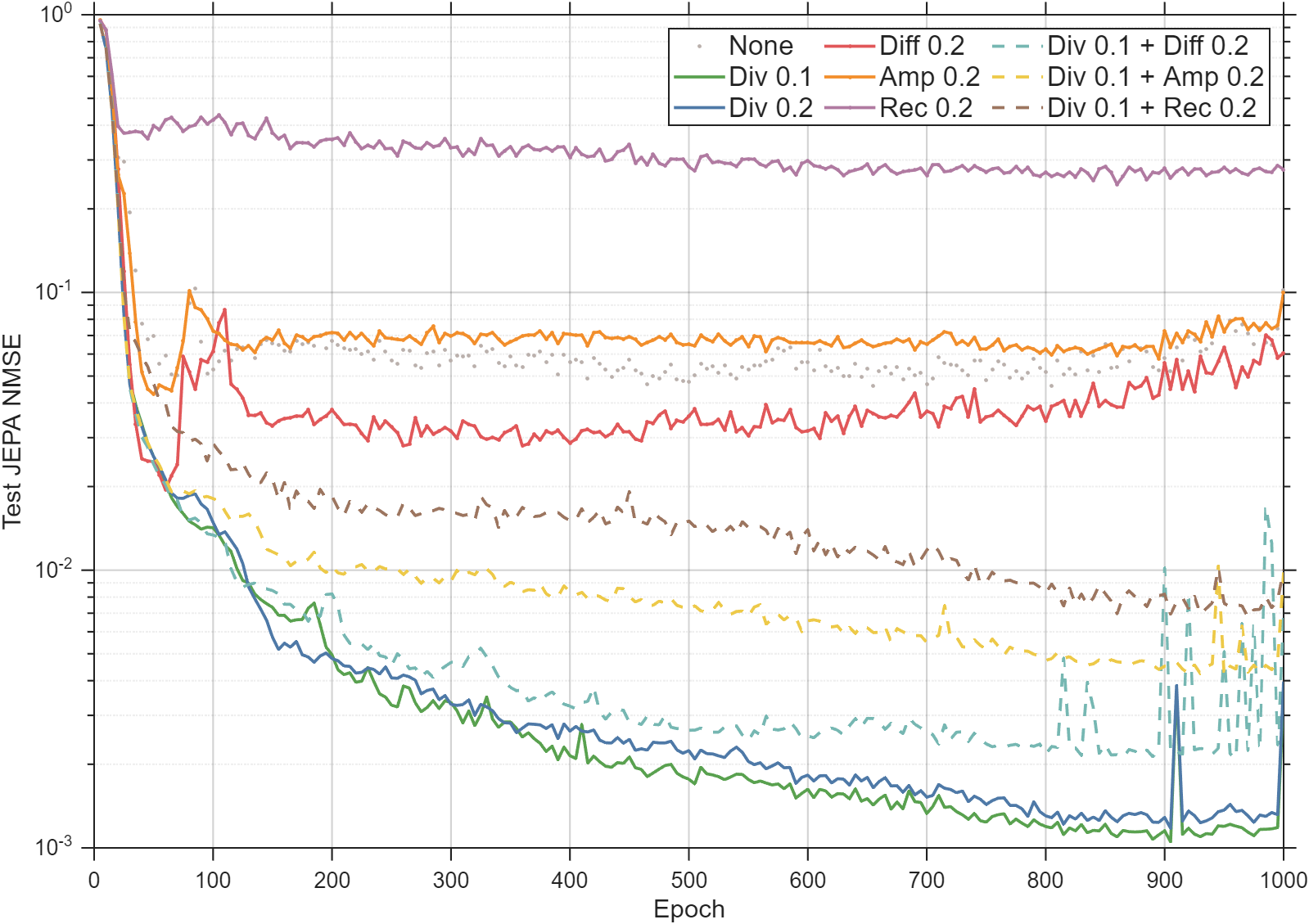}
  \caption{Single-scale objective ablation on \(32\times16\) observations, measured by test JEPA prediction NMSE.}
  \label{fig:jepa_results}
\end{figure}
\vspace{-0.15cm}
\subsection{Single-Scale Objective Ablation on \(32\times16\) Observations}

We first evaluate different pretraining objectives on the \(32\times16\) observation scale. The main comparison considers the JEPA prediction objective with different diversity weights, while CSI-domain grounding objectives are introduced as ablations. Although the model is optimized using the \(L_1\) latent prediction loss, we report the test latent-prediction NMSE, which provides a more interpretable measure of prediction quality. When a grounding objective is enabled, we set \(\lambda_{\rm g}=0.2\), as selected in preliminary experiments.

Fig.~\ref{fig:jepa_results} shows that diversity regularization is the main factor allowing accurate latent prediction. Without diversity
regularization, the JEPA objective remains substantially higher NMSE, and direct CSI recovery is particularly detrimental. In contrast, both diversity-only settings reduce the final prediction error by more than one order of magnitude, with \(\lambda_{\rm div}=0.1\) slightly outperforming \(\lambda_{\rm div}=0.2\).

Adding CSI-domain grounding on top of the diversity objective does not further improve latent prediction. Among the three grounding variants, the final NMSE increases in the order of diffusion grounding, amplitude recovery, and direct CSI recovery. This ordering follows the amount of channel-domain detail imposed on the latent representation: the more coefficient-level information the latent is required to preserve, the harder it becomes to predict from historical context. We therefore use the JEPA-diversity objective with \(\lambda_{\rm div}=0.1\) as the default single-scale pretraining objective.
\begin{figure}[t]
  \centering
  \includegraphics[width=0.65\linewidth]{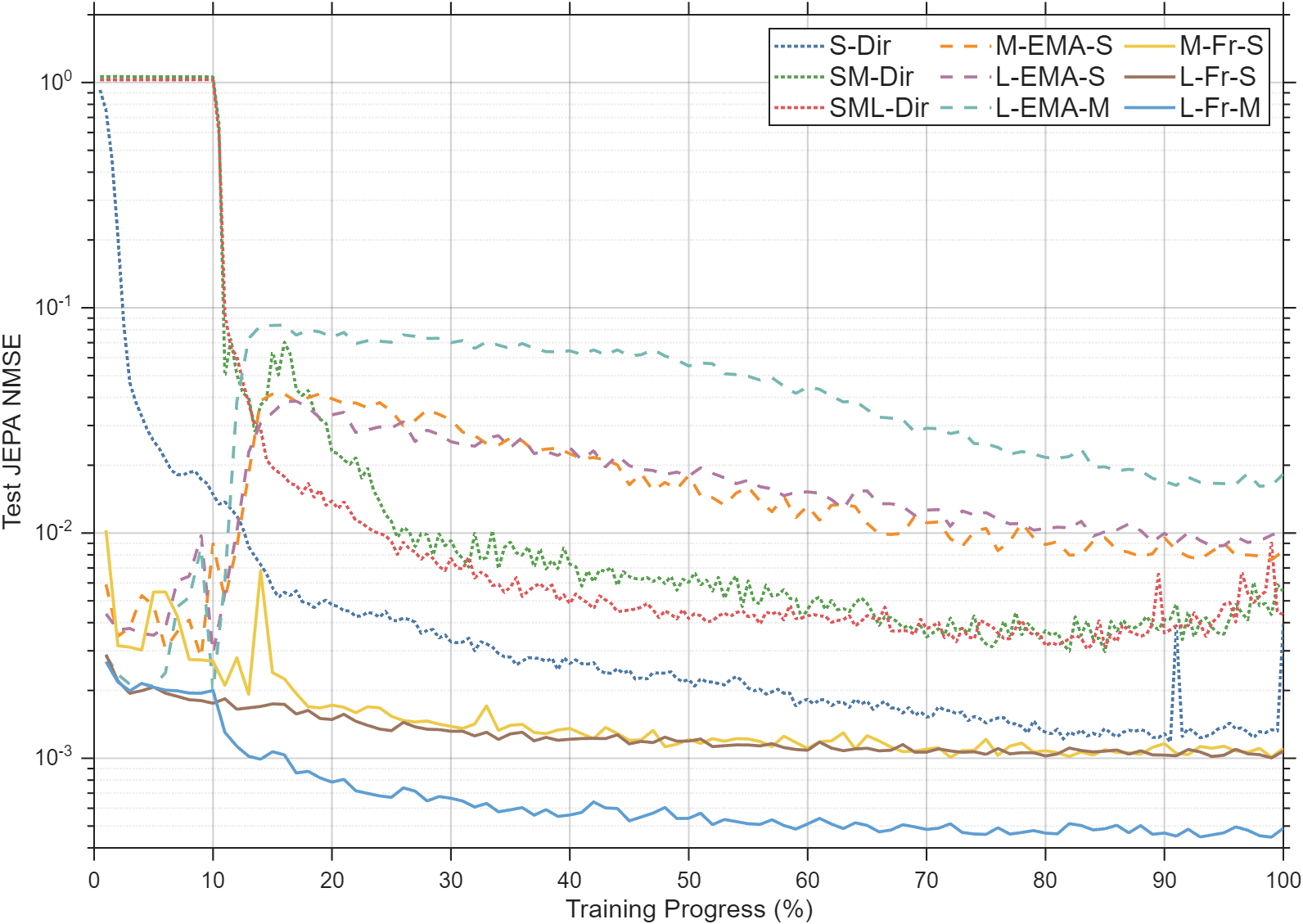}
  \caption{Test JEPA prediction NMSE under different multi-scale training strategies.}
  \label{fig:strategy}
\vspace{-0.7cm}
\end{figure}
\begin{figure}[t]
  \centering
  \includegraphics[width=0.75\linewidth]{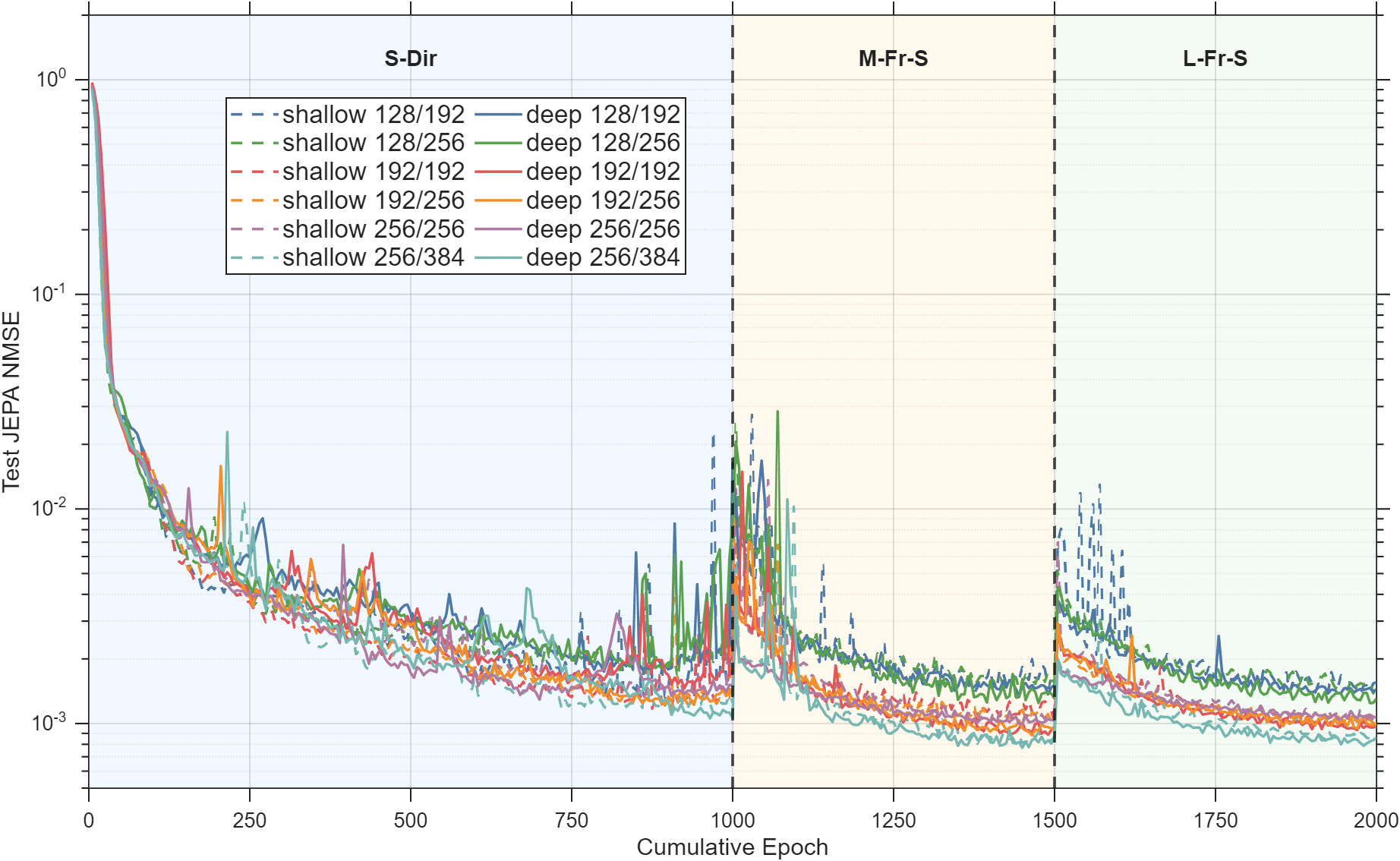}
  \caption{Topology comparison under the staged \(\mathrm{S\text{-}Dir}\rightarrow\mathrm{M\text{-}Fr\text{-}S}
\rightarrow\mathrm{L\text{-}Fr\text{-}S}\) training sequence. Solid and dashed curves denote deep and shallow backbones, respectively.}
  \label{fig:JEPA_multiscale}
\vspace{-0.7cm}
\end{figure}
\vspace{-0.5cm}
\subsection{Multi-Scale Training Strategy}

We denote the \(32\times16\), \(128\times64\), and \(1024\times64\) observation scales by ``S'', ``M'', and ``L'', respectively. The S scale is first trained from scratch as ``S-Dir''. At the M scale, we compare the joint training from scratch over S and M (``SM-Dir'') with incremental training from the pretrained S model using either a frozen reference branch (``M-Fr-S'') or an evolving branch supervised by an EMA teacher (``M-EMA-S''). At the L scale, the corresponding modes are direct three-scale training (``SML-Dir''), frozen-reference alignment from S or M (``L-Fr-S'' and ``L-Fr-M''), and EMA-reference alignment from S or M (``L-EMA-S'' and ``L-EMA-M'').

In the ``Dir'' modes, all included scales are initialized and trained jointly from scratch. In the ``Fr'' modes, the pretrained lower-scale branch is kept fixed and only the newly introduced scale is optimized. In the ``EMA'' modes, the old branch remains trainable while its teacher is updated by exponential moving average. For both incremental modes, the warmup stage optimizes only the new tokenizer using cross-scale alignment. The objective then gradually transitions from alignment to JEPA prediction according to the cosine schedule.

Fig.~\ref{fig:strategy} shows that keeping the pretrained reference branch fixed is consistently more effective than either joint multi-scale training from scratch or an evolving EMA reference. The frozen-reference variants converge rapidly and reach lower, more stable latent-prediction NMSEs, whereas the direct multi-scale and EMA variants retain noticeably larger residual errors. The EMA variants also exhibit stronger transients after a new scale is introduced, indicating that reference drift makes cross-scale alignment less stable. Among the L-scale settings, ``L-Fr-M'' achieves the lowest latent-prediction NMSE, suggesting that a fixed and more closely matched M-scale reference facilitates the introduction of the largest tokenizer. These results support frozen-reference incremental alignment as the preferred
scale-expansion strategy.
\vspace{-0.5cm}
\subsection{Effects of Architecture Scaling}
Fig.~\ref{fig:JEPA_multiscale} compares different latent/predictor dimension pairs and shallow or deep backbones under the common staged sequence \(\text{S-Dir}\rightarrow\text{M-Fr-S}\rightarrow\text{L-Fr-S}.\) The introduction of each new observation scale causes a short transient increase in prediction error, after which all configurations recover and continue to improve. This behavior confirms that the staged alignment procedure can preserve the previously learned latent field while incorporating a new tokenizer.

Higher-capacity variants attain the lowest final JEPA NMSE, with the deep \(256/384\) configuration giving the best final result. However, the improvements are not monotonic across all dimension pairs, and the differences among most configurations are considerably smaller than those caused by the training strategy in Fig.~\ref{fig:strategy}. Therefore, increasing latent or backbone capacity is beneficial only up to a point, while stable cross-scale optimization remains the more important design factor.

\vspace{-0.35cm}
\section{Downstream Channel Reconstruction and Task-Level Performance}
\label{sec:downstream-results}
\vspace{-0.1cm}
\subsection{Baseline and Criterion Methods}

To isolate the contributions of latent prediction, current pilot observations, and the two-stage training scheme, we compare the proposed FWM with the following four reference methods.
\begin{itemize}
  \setlength{\itemsep}{1pt}
  \setlength{\parskip}{0pt}
  \setlength{\parsep}{0pt}

  % \item \textbf{Simple Prediction (SP):}
  % SP is a non-JEPA channel-domain predictor that directly maps historical CSI observations to the current target channel. It does not use any current pilot observation and therefore represents a prediction-only baseline.

  \item \textbf{Simple Estimation (SE):}
  SE reconstructs the target channel solely from the current sparse and noisy pilot-based estimate. Historical observations and the predicted latent prior are both removed, making SE an estimation-only baseline.

  \item \textbf{Direct Training (DT):}
  DT uses the same input information and reconstruction architecture as FWM, but removes the JEPA pretraining stage. Instead, the complete network is initialized from scratch and optimized end-to-end for 500 epochs using only the final channel reconstruction objective. It therefore evaluates the benefit of the proposed two-stage pretraining and downstream training scheme.

  \item \textbf{Current-Latent Reconstruction (CL):}
  CL is an oracle reference that replaces the predicted latent field with the latent representation extracted from the current channel and feeds it to the same downstream reconstruction network. Since the complete current channel is required to construct this latent, CL is not realizable in practice. Rather, it indicates the reconstruction performance under an ideal latent prediction.
\end{itemize}

The proposed method is denoted by \textbf{FWM}. Since frequency alignment across carrier bands (\textbf{FA}) is optional, FWM, DT, and CL are each evaluated in two forms, with and without FA. SP and SE are retained as single-purpose baselines. In particular, SE-FA transfers latent information extracted from the observed carrier bands to reconstruct the missing carrier bands. Depending on the definition of the source and target carrier bands, this formulation is structurally similar to FDD cross-band channel prediction~\cite{liu2021fire}.\footnote{The SE baseline with frequency alignment has the same problem formulation as \cite{liu2021fire}. We only adjust the network topology to fit our data structure.}

We do not report pure-prediction baselines because none converged in our experiments, including \cite{xiao2025odeformer,cheng2026csi} and the prediction-only counterpart of the proposed method. This behavior is partly attributable to the random spatial perturbations introduced in our dataset, which make phase-level CSI prediction highly unreliable.

For FA settings, the availability of current carrier-band observations is randomized during training. For each sample, we first draw \(N_{\rm miss}\sim\mathcal{U}\{1,\ldots,7\}\), and then randomly select \(N_{\rm miss}\) out of the eight carrier bands as missing targets. The remaining \(N_{\rm vis}=8-N_{\rm miss}\) observed carrier bands provide current pilot observations. During testing, \(N_{\rm vis}\) is fixed for each evaluation condition to examine performance under different amounts of available frequency-domain information.

All methods are evaluated by their channel reconstruction quality. For the non-FA setting, we additionally transmit uncoded 16-QAM symbols on the non-pilot resource elements, use the reconstructed CSI for equalization and detection, and report the symbol error rate (SER).

For the FA setting, we further evaluate whether the reconstructed CSI can support single-stream transmit beamforming on the missing carrier bands. After OFDM demodulation, each OFDM subcarrier is modeled as a narrowband MISO channel. Let \(\widehat{\mathbf H}_{n,T}\in\mathbb{C}^{1024\times 64}\) denote the reconstructed channel of the carrier band indexed by \(n\), where \(\widehat{\mathbf h}_{n,k,T}^{H}\) is its \(k\)-th OFDM-subcarrier channel. We assign one unit-norm beamforming vector to each carrier band and share it across all OFDM subcarriers within that carrier band. Thus, the beamformer is designed from the reconstructed CSI by maximizing the average received power over the OFDM subcarriers of the carrier band:
\begin{equation}\footnotesize
\widehat{\mathbf w}_{n,T}=\arg\max_{\|\mathbf w\|_2=1}\frac{1}{N_{\rm sc}}\sum_{k=1}^{N_{\rm sc}}\left|\widehat{\mathbf h}_{n,k,T}^{H}\mathbf w\right|^2=\mathbf v_{\max}\left(\widehat{\mathbf H}_{n,T}^{H}\widehat{\mathbf H}_{n,T}\right),
\end{equation}
where \(\mathbf v_{\max}(\cdot)\) denotes the normalized principal eigenvector. This band-wise design accounts for the frequency-selective OFDM channel while avoiding an independent beamformer for every OFDM subcarrier.

\begin{figure}[t]
  \centering
  \includegraphics[width=0.85\linewidth]{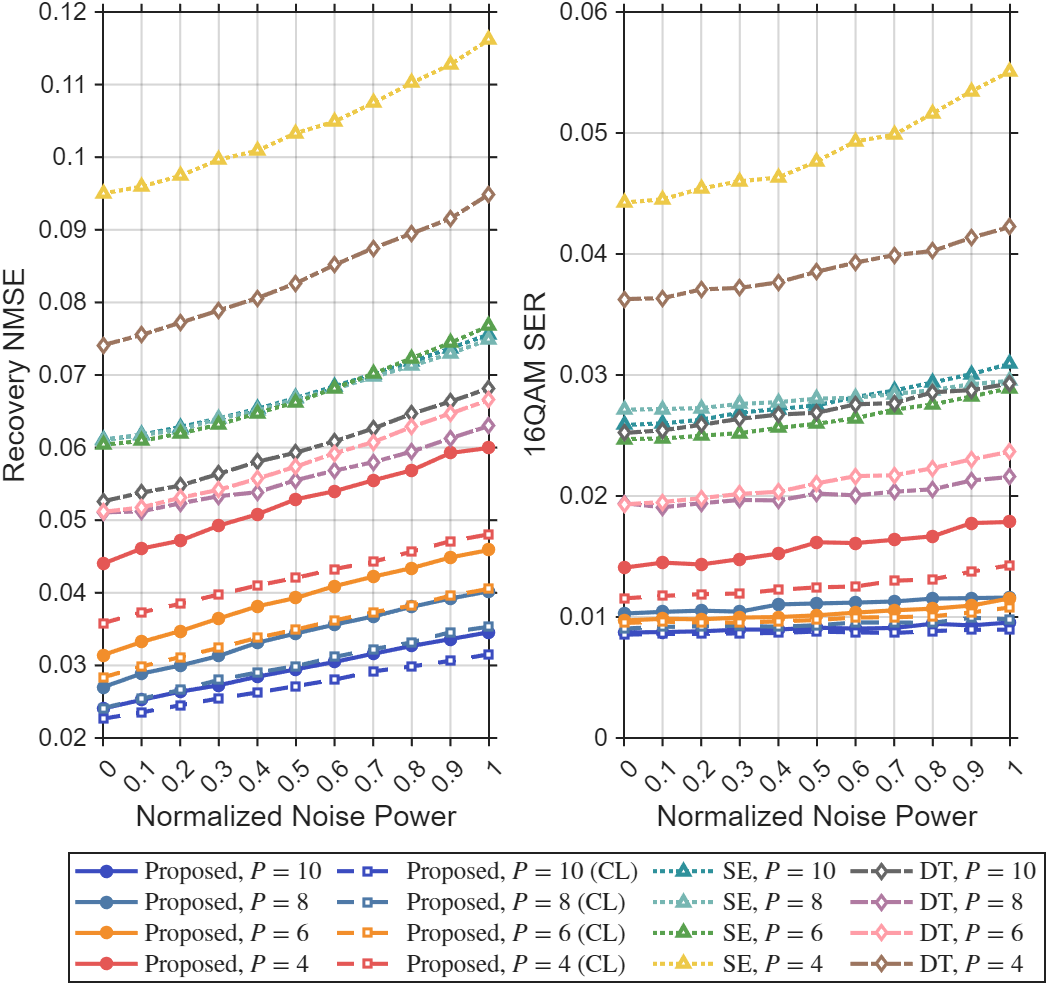}
  \caption{Single-band downstream performance without frequency alignment across carrier bands.}
  \label{fig:downstream_nocal}
\vspace{-0.3cm}
\end{figure}
\begin{figure}[t]
  \centering
  \includegraphics[width=0.8\linewidth]{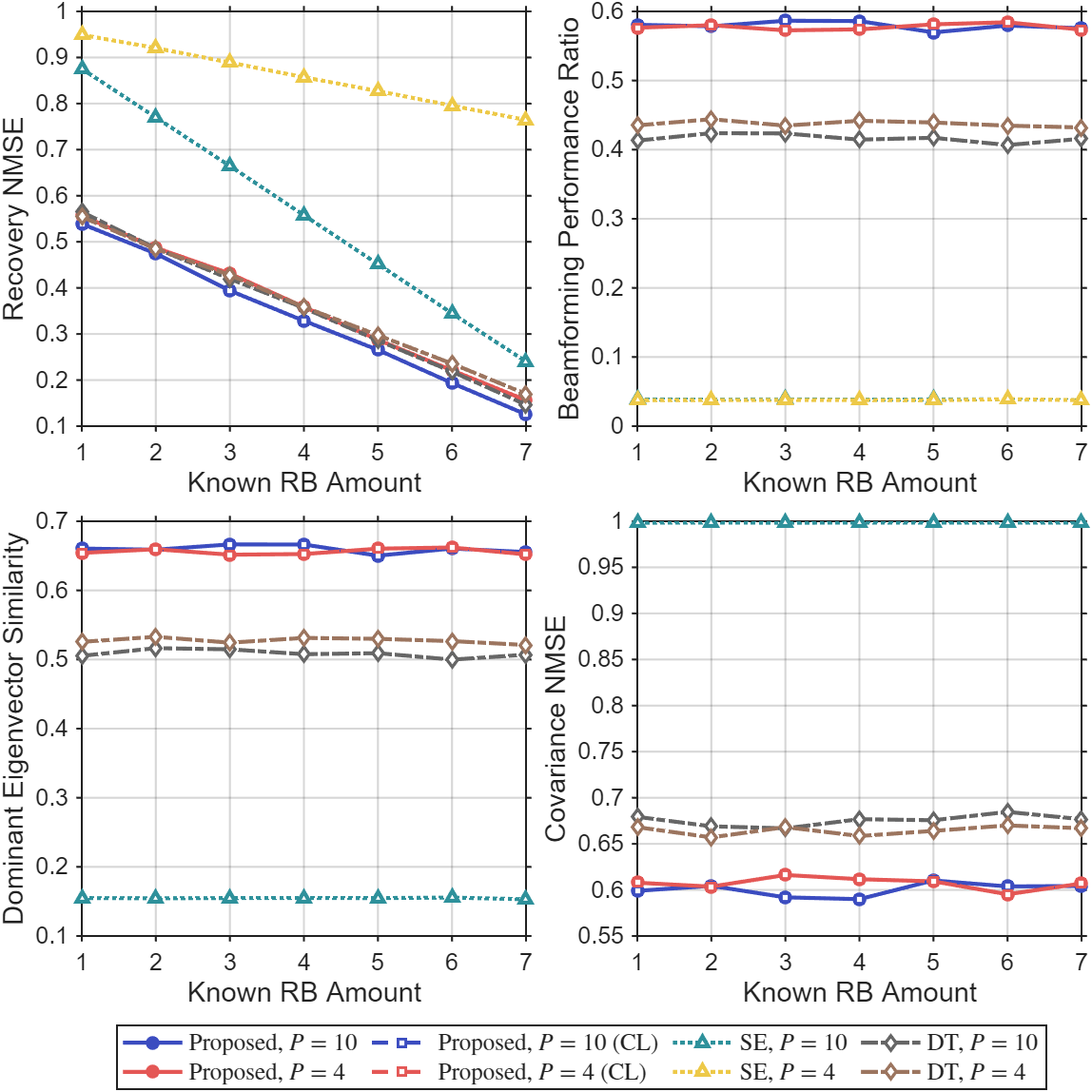}
  \caption{Frequency-alignment performance on the missing carrier bands for \(P=10\) and \(P=4\).}
  \label{fig:downstream_cal}
\vspace{-0.7cm}
\end{figure}
\begin{figure*}[t]
  \centering
  \subfloat[Grounding-objective ablation.\label{fig:down_ablation}]{%
    \includegraphics[width=0.32\textwidth]{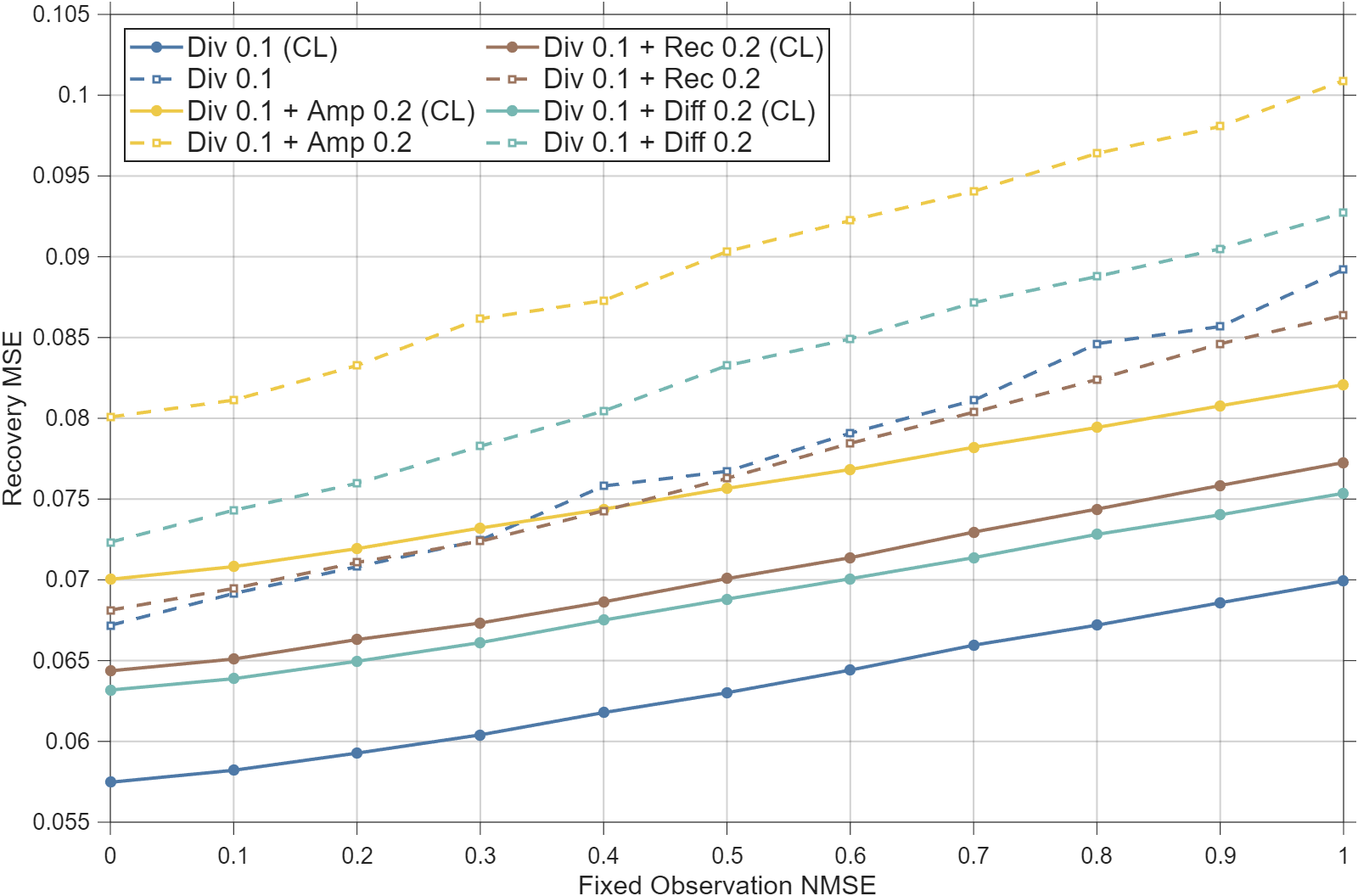}}
  \hfill
  \subfloat[Multi-scale training strategy.\label{fig:down_strategy}]{%
    \includegraphics[width=0.32\textwidth]{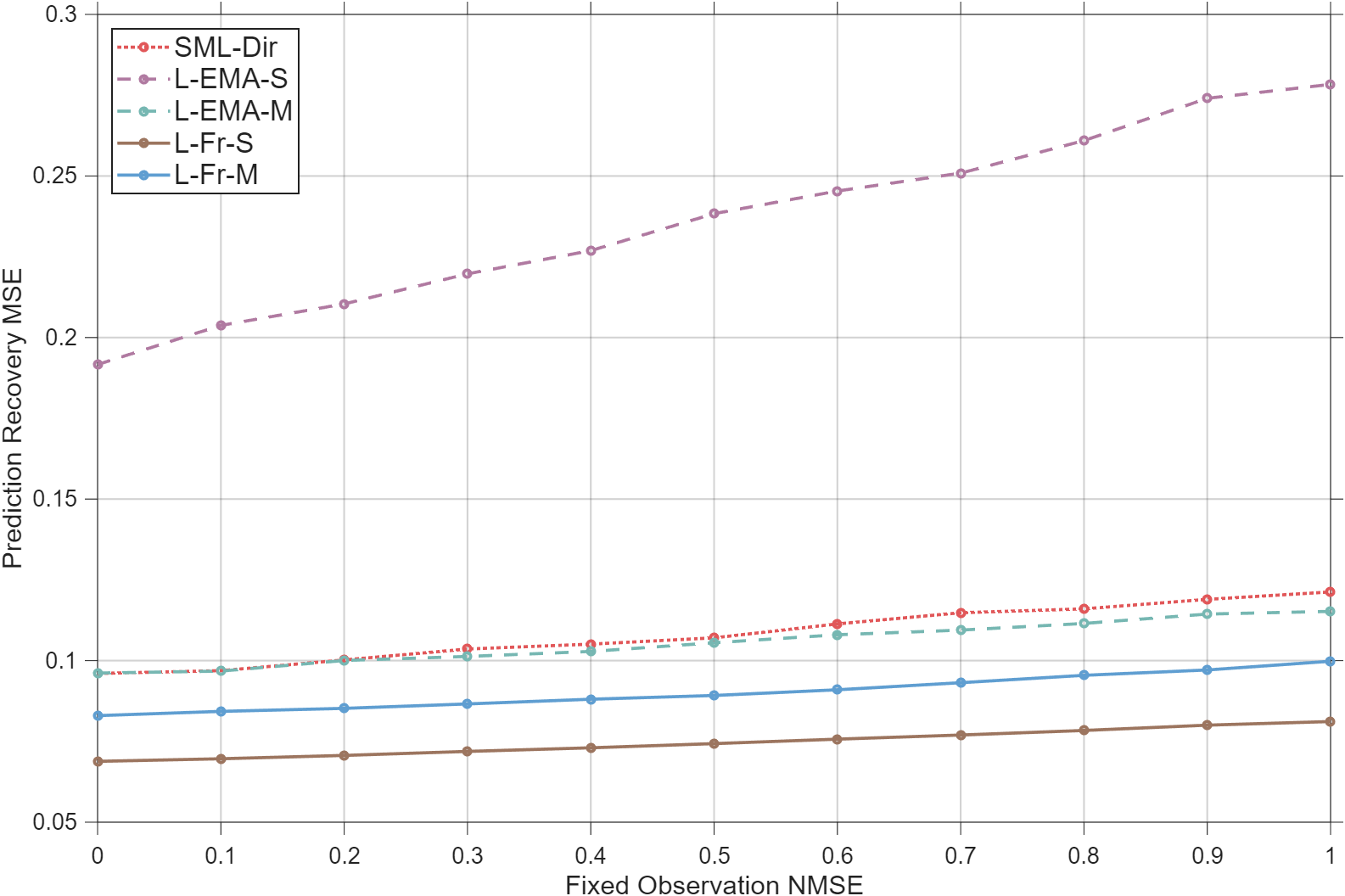}}
  \hfill
  \subfloat[Architecture scaling.\label{fig:down_topology}]{%
    \includegraphics[width=0.32\textwidth]{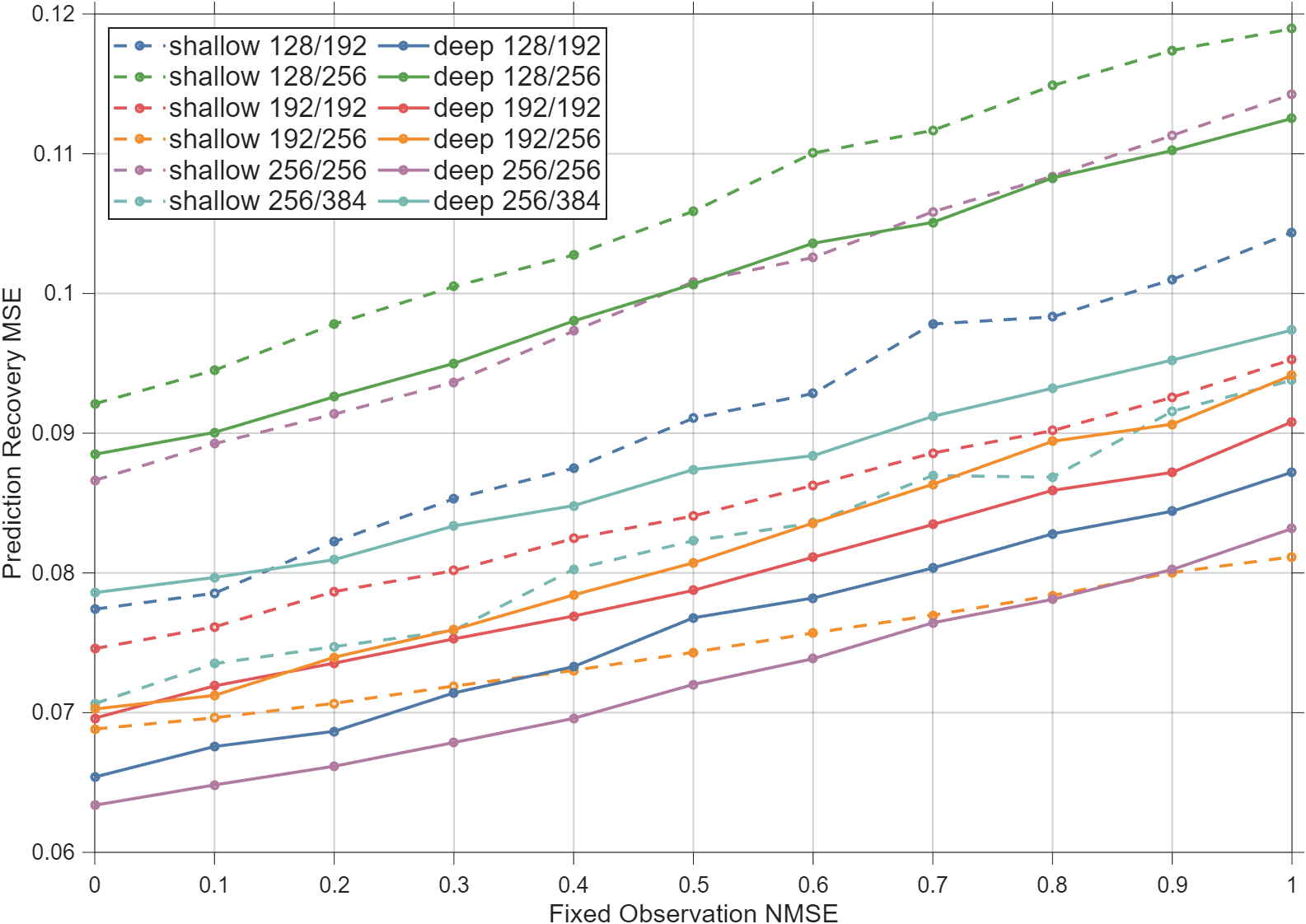}}
  \caption{Downstream reconstruction performance corresponding to the pretraining studies, evaluated versus fixed-observation NMSE.}
  \label{fig:downstream_results}
\end{figure*}
To quantify the spatial structure relevant to beamforming, we define the true and reconstructed covariance matrices of carrier band \(n\) as
\begin{equation}
\mathbf R_{n,T}
=
\frac{1}{N_{\rm sc}}\mathbf H_{n,T}^{H}\mathbf H_{n,T},
\,
\widehat{\mathbf R}_{n,T}
=
\frac{1}{N_{\rm sc}}\widehat{\mathbf H}_{n,T}^{H}\widehat{\mathbf H}_{n,T}.
\end{equation}
Let \(\mathbf w_{n,T}^{\star}=\mathbf v_{\max}(\mathbf R_{n,T})\) denote the perfect-CSI beamformer. The normalized beamforming performance ratio is
\begin{equation}
G_{n,T}
=
\frac{
\widehat{\mathbf w}_{n,T}^{H}
\mathbf R_{n,T}
\widehat{\mathbf w}_{n,T}
}{
\lambda_{1}(\mathbf R_{n,T})
},
\end{equation}
where \(G_{n,T}=1\) corresponds to perfect-CSI beamforming. We further report the dominant-eigenvector similarity and covariance NMSE,
\begin{equation}
S_{n,T}
=
\left|
(\mathbf w_{n,T}^{\star})^{H}
\widehat{\mathbf w}_{n,T}
\right|,
\mathrm{NMSE}_{\mathbf R,n,T}
=
\frac{
\left\|
\widehat{\mathbf R}_{n,T}-\mathbf R_{n,T}
\right\|_{F}^{2}
}{
\left\|
\mathbf R_{n,T}
\right\|_{F}^{2}
}.
\end{equation}
All three metrics are averaged over the evaluated missing carrier bands and test samples. Since \(\mathbf R_{n,T}\) is positive semidefinite, \(G_{n,T}\geq S_{n,T}^{2}\). Hence, preservation of the dominant spatial eigendirection directly limits the beamforming loss.

\vspace{-0.4cm}
\subsection{Single-Band Reconstruction and Symbol Detection}

We first evaluate independent reconstruction of each carrier band from its predicted latent prior and local sparse pilots. As shown in Fig.~\ref{fig:downstream_nocal}, FWM consistently outperforms SE and DT in both reconstruction NMSE and 16-QAM SER, with a larger advantage when fewer pilots are available. This behavior indicates that the predicted latent field supplies channel structure that becomes most valuable when instantaneous pilot evidence is sparse and noisy. The corresponding SER improvement further shows that this prior preserves information relevant to symbol detection rather than merely reducing coefficient error.

\subsection{Cross-Band Reconstruction and Beamforming}

We next activate frequency alignment: current pilots are available only in the observed carrier bands, while the missing carrier bands are reconstructed from these observations and the predicted latent field. Fig.~\ref{fig:downstream_cal} shows that FWM and DT have comparable coefficient-level reconstruction errors, whereas their beamforming performance differs substantially. FWM remains close to the current-latent reference and clearly outperforms DT and SE, with only weak dependence on the number of observed carrier bands. This indicates that the predicted latent prior already contains much of the spatial information required for beamforming.

The higher dominant-eigenvector similarity and lower covariance NMSE of FWM explain this gain. JEPA pretraining better preserves the dominant spatial eigendirection and covariance structure of the missing-band channel, which govern the beamforming Rayleigh quotient. The benefit therefore arises from spatial-subspace preservation rather than uniformly more accurate complex coefficients, explaining why similar reconstruction NMSE can lead to markedly different beamforming performance.

\subsection{Pretraining-Downstream Correspondence}

Fig.~\ref{fig:downstream_results} compares the main pretraining choices in downstream reconstruction. The grounding objectives in Fig.~\ref{fig:down_ablation} provide no consistent benefit over diversity-based JEPA pretraining, suggesting that retaining additional raw-CSI detail is unnecessary for forming a useful reconstruction prior. Frozen-reference incremental training remains preferable in Fig.~\ref{fig:down_strategy}, although latent-prediction NMSE does not perfectly determine downstream performance. Fig.~\ref{fig:down_topology} likewise shows no monotonic benefit from increasing model capacity. These results support selecting the pretraining strategy and architecture according to downstream utility rather than latent accuracy or scale alone.

\section{Conclusion and Future Directions}

In this paper, we proposed a FWM for wireless channel representation, prediction, and reconstruction. The central idea is to avoid direct coefficient-level prediction of raw CSI, whose phase-sensitive details can vary severely under small spatial perturbations and imperfect UE mobility information. Instead, the proposed FWM maps multi-resolution CSI observations across the considered carrier bands into a shared latent propagation-field space and performs prediction in this latent domain. We studied JEPA-based pretraining objectives, latent capacity, and incremental multi-scale alignment strategies, showing how tokenizer heads at different observation scales can be aligned to a common latent field. We further applied the learned latent field to direct single-band reconstruction and cross-band reconstruction with missing pilots, and evaluated both coefficient-level reconstruction and communication-level performance. The predicted latent prior improves symbol detection and, more importantly, provides a substantial band-wise beamforming gain even when its NMSE advantage over sufficiently trained end-to-end baselines is small or inconsistent. Since beamforming is governed primarily by the dominant spatial subspace rather than uniform coefficient accuracy, this result indicates that JEPA pretraining preserves task-relevant propagation structure beyond pointwise CSI fitting. Overall, FWM provides a step from channel approximation toward physics-aware field representation, while sparse current observations remain necessary for recovering instantaneous fine-grained channel details.

Several directions remain for future work. First, the present study mainly focuses on receiver-side channel reconstruction in a single ray-tracing scenario. A natural extension is to evaluate FWM across different propagation scenes, deployment geometries, carrier frequencies, and mobility patterns, and to extend the framework from channel reconstruction to multiple physical-layer tasks such as beam management, resource allocation, sensing, and localization. Second, the training data considered in this work are still relatively ideal, since the underlying full channel realizations are available to construct different observation scales and supervision targets. In practical systems, training data may be noisy, partially observed, biased by imperfect channel estimation, or missing at some scales and carrier bands. Future FWM training should therefore incorporate noisy and incomplete observations more explicitly. These extensions would move FWM closer to practical deployment as a general latent world-model backbone for heterogeneous wireless systems.
\appendices
\section{Tokenizer Architectures} \label{app:tokenizer}
\begin{figure}[t]
    \centering
    \includegraphics[width=0.85\linewidth]{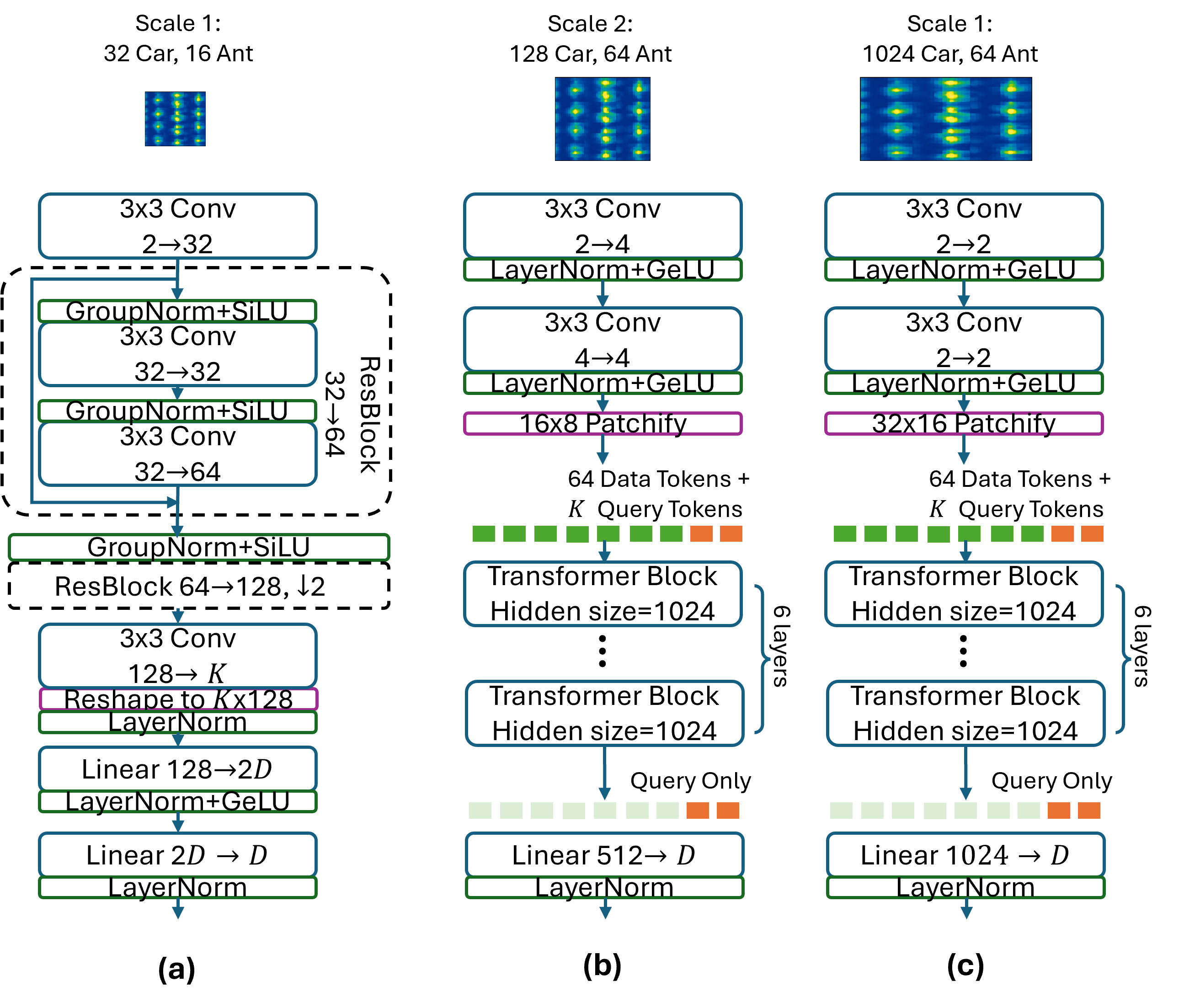}
\caption{An illustration of the architectures for tokenizers.}
    \label{fig:tokenizer}
\vspace{-0.3cm}
\end{figure}
At each observation scale, the low-frequency and high-frequency bands use two tokenizer heads with the same architecture but independent parameters. This keeps the processing pattern consistent across frequency groups, while allowing different spectral regions to adapt to their own channel statistics. Unless otherwise specified, each tokenizer outputs \(K=2\) latent tokens for each time-band cell, with token dimension \(D=128\). Therefore, the main difference among tokenizers lies in how the input spatial dimensions are compressed before token formation.

Fig.~\ref{fig:tokenizer} summarizes the tokenizer architectures. For the \(32\times16\) scale, we use a CNN tokenizer. The real/imaginary two-channel input is first embedded by a convolutional stem, then compressed by residual downsampling blocks, and finally converted into token maps. Each token map is flattened and projected to the target latent dimension by a lightweight MLP.

For the \(128\times64\) and \(1024\times64\) scales, we use ViT-style tokenizers. A shallow convolutional stem first extracts local features, which are then partitioned into non-overlapping patches. The patch sizes are \(16\times8\) and \(32\times16\) for the \(128\times64\) and \(1024\times64\) inputs, respectively, producing 64 and 128 patch tokens. After adding positional embeddings, a small set of learnable query tokens is concatenated with the patch tokens and processed by transformer blocks. Only the output query tokens are retained and projected to the final latent dimension. A layer normalization layer is applied at the output of all tokenizer heads to stabilize the latent scale.

\section{Prediction Backbone Architectures} \label{app:pred}
\begin{figure}[t]
    \centering
    \includegraphics[width=0.85\linewidth]{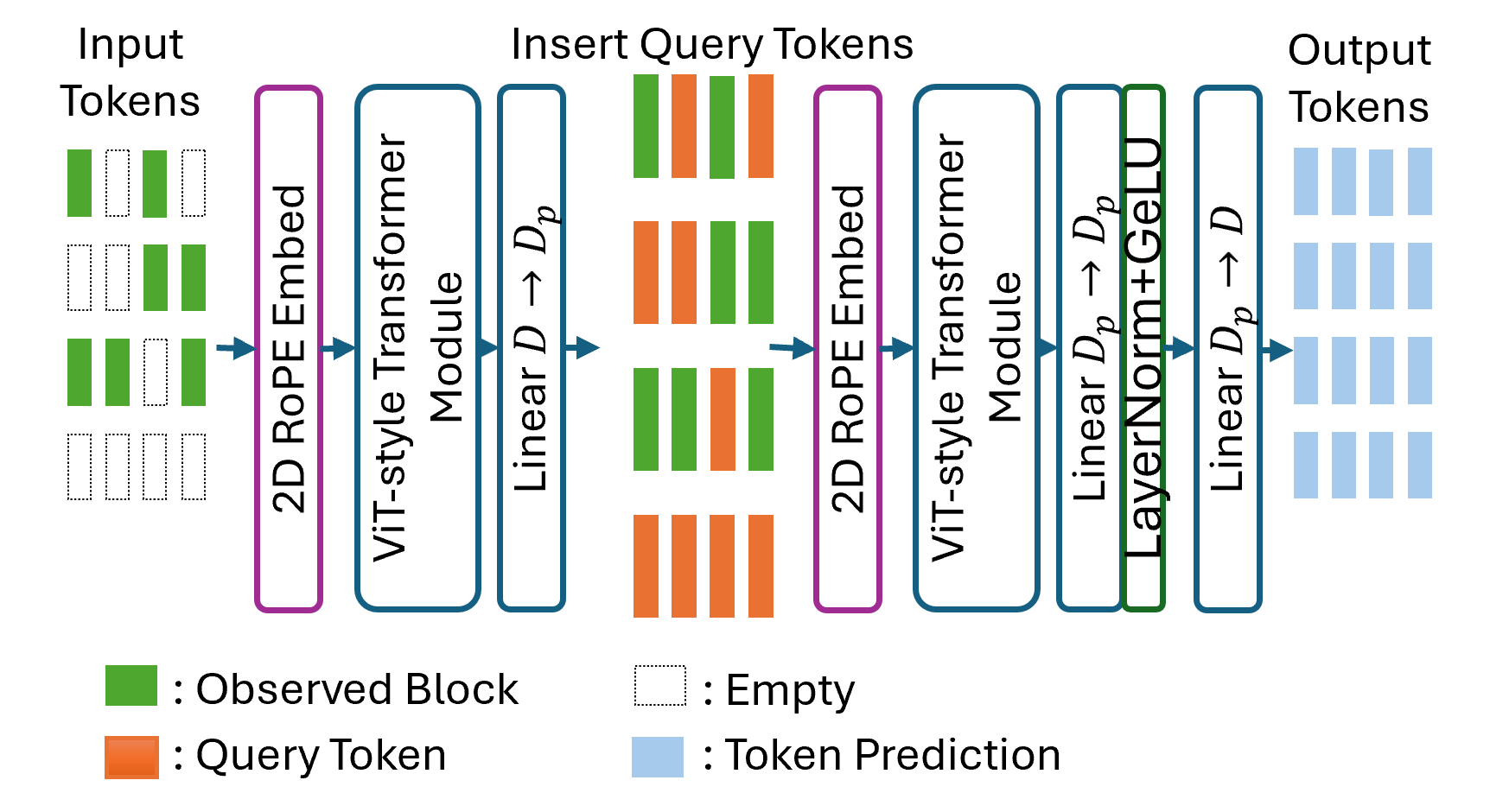}
    \caption{The structure of the encoder and predictor for JEPA.}
    \label{fig:pred}
\end{figure}
As illustrated in Fig.~\ref{fig:pred}, the JEPA encoder and predictor operate on tokenized time--band cells rather than on spatial channel grids. Both modules are implemented as ViT-style networks. Positional information is injected through two-dimensional rotary embeddings over the time and band indices, together with a learnable embedding for the token index within each cell.
The context encoder processes only visible latent tokens and produces contextualized field representations. The predictor then takes these context representations together with learnable query tokens for masked cells and predicts the corresponding target latents. Linear projections are used only to match the encoder, predictor, and tokenizer latent dimensions.
We consider two backbone sizes. The ``shallow'' configuration uses a 4-layer, 8-head ViT encoder and a 2-layer, 4-head ViT predictor. The ``deep'' configuration uses a 6-layer, 8-head ViT encoder and a 4-layer, 8-head ViT predictor.
\vspace{-0.4cm}
\section{Grounding-Head Architectures} \label{app:aux}
For the CSI-domain grounding ablations, we use lightweight auxiliary heads attached to the JEPA latent tokens. These heads are used only during the corresponding ablation experiments and are not part of the default FWM pretraining objective.

The raw CSI recovery (``Rec'') and amplitude-only recovery (``Amp'') ablations share the same CNN recovery decoder. The latent tokens are first projected back to a compact feature map and then decoded to the channel domain through a small convolutional decoder. The ``Rec'' task supervises the recovered complex-valued channel tensor, whereas the ``Amp'' task applies the loss only to the recovered channel amplitude. Therefore, the two ablations use the same architecture and differ only in the supervision target. The decoder has a structure symmetric to that of the \(32\times16\) tokenizer shown in Fig.~\ref{fig:tokenizer}(a).

The ``Diff'' ablation uses a lightweight conditional DiT-style head, following a standard DDPM design with 1000-step noise schedule. The noisy channel sample, diffusion step, and JEPA latent tokens are embedded to a common hidden space, and the head predicts the injected noise through a shallow transformer stack. In our implementation, the hidden dimension is 128, the depth is 3, the number of attention heads is 8, and the MLP ratio is 4.0. All auxiliary heads are instantiated separately for the low-frequency and high-frequency branches.

\section{Reconstruction Head Architectures} \label{app:recon}
\begin{figure}[t]
    \centering
    \includegraphics[width=0.9\linewidth]{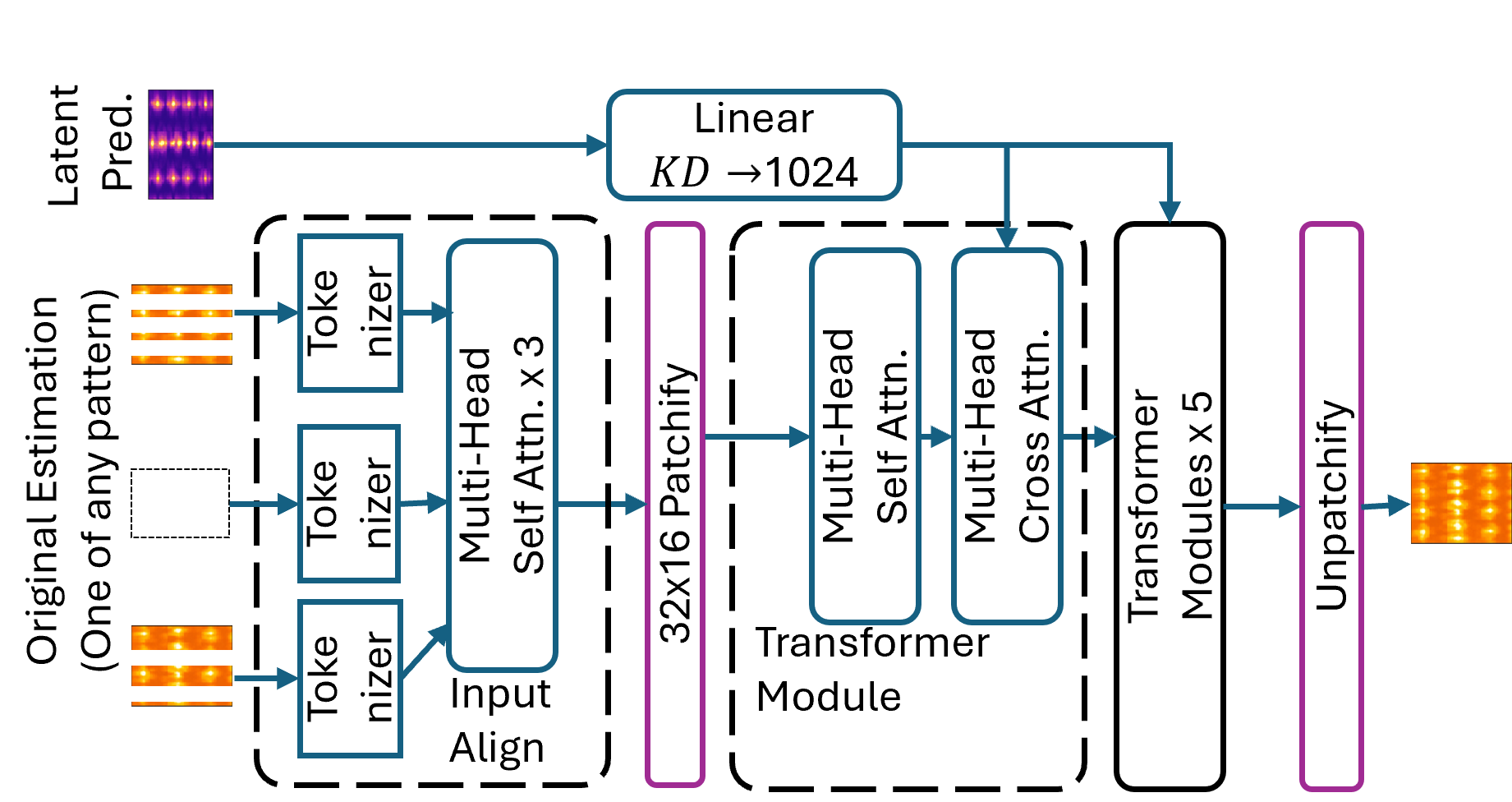}
    \caption{The structure of the reconstruction head for the downstream task.}
    \label{fig:recon}
\vspace{-0.5cm}
\end{figure}
For downstream channel reconstruction, historical observations are first processed by the pretrained FWM to predict the current latent representation. Depending on the pretrained backbone, this latent prior may come from the \(32\times16\), \(128\times64\), or \(1024\times64\) JEPA model. In all cases, the downstream interface is unified as a sequence of per-band latent vectors. The overall model is shown in Fig.~\ref{fig:recon}.

First, the latent alignment module refines the predicted latent vectors using the available current pilot-based observations from observed carrier bands. It consists of a fusion head followed by self-attention blocks over the band dimension. The fusion head shares the structure of the \(1024\times64\) tokenizer but is initialized and trained independently. The output is a calibrated latent condition that can be used by the reconstruction head for both observed and missing carrier bands. This module is optional; when it is omitted, we directly use the predicted latent as the calibrated output.

The downstream channel reconstruction head operates on the full \(1024\times64\) channel grid. The sparse current estimate is partitioned into non-overlapping \(32\times16\) patches, producing a \(64\times4\) patch grid with 256 patch tokens. A convolutional patch embedding maps each patch to a hidden representation of dimension 1024. The latent condition is projected to the same hidden dimension and injected through cross-attention. After 6 transformer modules, each patch token is projected back to a \(32\times16\) channel patch, and all patches are reassembled into the recovered \(1024\times64\) channel estimate.
\printbibliography
\end{document}